\documentclass[draft]{agujournal2019}
\usepackage{url} 
\usepackage{lineno}
\usepackage[inline]{trackchanges} 
\usepackage{soul}
\usepackage{blindtext}
\usepackage{wrapfig}
\usepackage{amsmath}
\usepackage{siunitx}
\usepackage{booktabs}
\usepackage{graphicx}
\usepackage{array}
\usepackage{adjustbox}
\usepackage{nameref}  
\usepackage{hyperref}
\usepackage{natbib}
\usepackage{amssymb}
 
\providecommand{\apjs}{ApJS}
\providecommand{\aap}{A\&A}

\let\oldmaketitle\maketitle
\renewcommand{\maketitle}{\oldmaketitle\setcounter{footnote}{0}} 
\draftfalse 

\journalname{Journal of Geophysical Research}

\begin{document}
\title{{\centering Comparative Analysis of 10 - 50\,MeV Solar Proton Events at Lagrange Point 1 and \\the Geostationary Orbit}}
 
\authors{Aatiya Ali, Viacheslav Sadykov}

\affiliation{1}{Georgia State University} 

\correspondingauthor{Aatiya Ali}{aali87@gsu.edu}

\begin{keypoints}
\item We apply machine-learning–based correction methods to mitigate previously identified flux contamination in GOES proton data.

 \item Random Forest performed most reliably, significantly reducing contamination.
 
 \item Incorporating geomagnetic indices into the model inputs resulted in modest gains in ${R}^{2}$ and reductions in RMSE.
 
 \item This study demonstrates the potential of using this approach to correct GOES proton measurements. Further refinement and optimization of the models could improve solar energetic particle forecasting accuracy and contribute to the success of future space exploration missions.
\end{keypoints}

\begin{abstract}
Solar proton events (SPEs) pose radiation hazards, disrupt technology, and impact operations on Earth and in space, making continuous monitoring essential. We compare 10 - 50\,MeV proton flux measurements from SOHO/EPHIN at Lagrange Point~1 (L1) with those from NOAA/GOES in geostationary orbit (GEO) during Solar Cycle~23 and most of Cycle~24. We identify 83 $\geq$10\,pfu SPEs observed at both locations and classify them into S1--S4 categories (comparable to NOAA’s solar radiation storm scales). EPHIN detected earlier onsets and longer durations across all categories, along with earlier peaks and ends for S1--S3; while GOES recorded slightly earlier peak/end times for S4. Median timing offsets (EPHIN relative to GOES) were $-20\pm50$\,min (S1 onsets), $-1.00\pm1.42$\,hr (S1 peaks), $-1.08\pm2.21$\,hr (S1 ends), with similar trends for S2--S3 and near-simultaneity for S4 (peaks $\sim-0.17\pm1.62$\,hr; ends $\sim+0.04\pm3.33$\,hr). Flux comparisons show that EPHIN measurements modestly exceed GOES for S1 (median ratios $\sim1.11$ for peaks and $\sim1.06$ for fluence) and are lower than GOES for stronger events (peaks $\sim0.97\pm{}0.29$, $0.84\pm{}0.21$, ; fluence $\sim0.84\pm{}0.16$, $0.75\pm{}0.16$ for S2--S3). The EPHIN-to-GOES peak flux and fluence ratios reach $0.16\pm{}0.03$ and $0.29\pm{}0.07$, correspondingly, for S4 events, originating from strong contamination of lower-energy GOES channels. Correlation analyses show no significant EPHIN-to-GOES flux dependence on geomagnetic indices, field strength, or spacecraft position, suggesting minimal near-Earth modulation of $\geq$10\,MeV proton access at GEO by near-Earth conditions. These results highlight systematic differences in how SPEs manifest at L1 versus GEO and offer practical guidance for forecasting beyond Earth’s magnetosphere, supporting mission planning for near-Earth and cislunar exploration, including Artemis.
\end{abstract}

\section{Introduction}   
 
    Solar energetic particle (SEP) events are enhanced fluxes of high-energy particles (mainly protons, electrons, and heavy ions) originating from the Sun. These particles span a wide energy spectrum, from keVs to multiple GeVs, as they traverse the heliosphere \citep{sep_energies_anastasiadis}. A major subclass, known as solar proton events (SPEs), is typically defined as excesses of protons with energies exceeding \(10 \, \text{MeV}\) and fluxes surpassing \(10 \, \text{particle flux units}\) (pfu), where \( 1 \mathrm{pfu}=1 \mathrm{particle}\cdot{}\mathrm{cm}^{-2}\mathrm{s}^{-1}\mathrm{sr}^{-1} \). The high-energy protons in these events can pose substantial risks to astronauts, satellites, navigation-communication systems, and other vulnerable technological assets. When SEPs impact spacecraft, they can interfere with electronic systems, leading to malfunctions in spacecraft calibration or even complete failure of crucial equipment. In addition, high-energy electrons can intensify operational risks by penetrating satellite shielding and compromising system integrity \citep{lsp,electrons}. Raising radiation levels, extreme SEP events can affect planetary environments and cause concern for human space missions beyond Earth's magnetospheric protection, posing health risks to astronauts and contributing to long-term physiological challenges such as radiation sickness, cancer, nervous system damage, and tissue degradation \citep{health_risk_onorato,sickness_lee}.  
    
    Addressing these concerns is essential, especially as interest grows in space exploration and commercial space travel. In particular, lunar exploration is uniquely affected by SEPs. Positioned at an average distance of $\sim$380,000 km from Earth, the Moon remains outside Earth’s magnetosphere for most of its orbit, exposing its surface to radiation from solar wind ions, high-energy SEPs, and galactic cosmic rays (GCRs) with minimal attenuation \citep{radiation_dandouras}. Using observations from the Time History of Events and Macroscale Interactions during Substorms (THEMIS) and the Acceleration, Reconnection, Turbulence, and Electrodynamics of the Moon's Interaction with the Sun (ARTEMIS) mission, \citet{main_moon_liuzzo} detected hazardous charged particles even when the Moon was embedded \textit{within} the magnetosphere. In contrast, on Earth, the magnetosphere can act as a shield, scattering low-energy particles back into space and limiting their access to the geostationary and low-Earth orbits \citep{bsphere_deflecion_liu}. Therefore, it becomes essential to understand the commonalities and differences between SEP events observed in two distinct environments: those without a protective magnetosphere (such as the Moon or the Lagrange point 1, L1) and those within Earth's magnetosphere. This knowledge is critical for predicting astronaut exposure risks, assessing long-term effects of space weather on the lunar environment, and understanding the mechanisms by which SEPs reach the Moon \citep{main_moon_liuzzo}. The sporadic and intense nature of SPEs with significant variability in flux and energy can lead to concerning radiation doses if adequate shielding is unavailable. During major SPEs, the fluence of protons with energies \(>30 \, \text{MeV}\) can surpass \(10^{10} \, \text{cm}^{-2}\) in several hours or days, posing operational constraints on human missions by depositing large radiation doses for crew and equipment that may not be adequately shielded \citep{exploration_hu}. To address this, numerous studies \citep[e.g.,][]{shelter_pham,shelter_martha} have focused on conceptualizing habitat designs and shelter requirements for the lunar surface during periods of high radiation exposure. These efforts examine factors such as material composition, shelter architecture, and environmental conditions on the Moon, exploring different strategies for effective shielding to safeguard astronauts during intense radiation events.
     
    Since the 1970s, near-Earth particle radiation measurements detail proton fluxes across energy levels critical for spacecraft shielding \citep{instrument_jiggens}. The Geostationary Operational Environmental Satellite (GOES) series, launched by the National Oceanic \& Atmospheric Administration (NOAA) and located at the geostationary orbit (GEO), have been instrumental in analyzing Earth-directed SEP events. In this study, we utilize proton flux data collected by the GOES series during the 1996-2015 time period. However, we also consider another instrument, the Electron, Proton, and Helium INstrument \citep[EPHIN,][]{ephin_eg1_kuhl,ephin_noise_kuhl} aboard the Solar and Heliospheric Observatory \citep[SOHO,][]{Domingo1995SoPh..162....1D}, located at L1 beyond Earth's magnetosphere. While not exactly at the location of the Moon, EPHIN measurements more closely represent the lunar environment than GOES data, as they sample SEPs beyond Earth’s magnetosphere while remaining in near-Earth space. This work specifically focuses on 10 - 50\,MeV proton fluxes as the fluxes in this energy range can be compared more or less explicitly between two instruments. Although the 10 – 50 MeV channel is not the $\geq$10 MeV integral flux that defines NOAA’s solar radiation storm scales (S-scales)\footnote{\url{https://www.swpc.noaa.gov/noaa-scales-explanation}}, it typically accounts for the bulk of the proton flux during weaker SEP events and dominates the peak fluxes in stronger events. The S-scale is a communication tool that classifies SPE severity using the GOES $\geq$10 MeV integral proton flux and relates event intensity to expected impacts- satellite operations, high-frequency communications, navigation, and radiation risk to astronauts and airline crew/passengers \citep{bain_2021}. Thresholds are: S1 = 10 pfu, S2 = $10^{2}$ pfu, S3 = $10^{3}$ pfu, S4 = $10^{4}$ pfu, and S5 = $10^{5}$ pfu. We use the same thresholds, but apply them to the flux of 10-50\,MeV protons only according to GOES measurements.
    
    Although previous studies have compared SEP properties both within and outside the magnetosphere \citep[e.g.,][]{arrival_direction_filwett}, to the best of our knowledge, a statistical investigation specifically for relatively energetic 10 - 50\,MeV protons at L1 has not been conducted before. In this work, we analyze SPEs in terms of their peak flux strength within the classes that are comparable (but not identical) to the NOAA S-scale classifications, and compare event properties observed at L1 with those detected at GEO. This allows us to identify systematic differences across event magnitudes and evaluate how the magnetosphere or instrumental effects influence SEP signatures measured at GEO. Evaluating SPEs through the strength-dependent S-scale framework is critical because it provides an operationally relevant measure of space weather impacts as SEPs propagate to GEO. Ultimately, this analysis not only enhances our understanding of particle transport and precipitation processes but also supports the development of improved prediction models essential for safeguarding astronaut health and ensuring mission reliability in lunar and cislunar environments \citep{guo_2023}.

\subsection{Scope of this Work} \label{scope} 

    The primary goal of this work is to compare properties of SPEs in the 10 - 50\,MeV energy range as observed at the L1 point and at GEO during Solar Cycles (SCs) 23 and 24 (partially). This study is motivated by two objectives. First, because characteristics of $\geq$10 MeV SPEs, such as onset times, peak fluxes, and fluences, are central to operational forecasting, comparing their low-energy components at L1 and GEO can reveal how location, environment, or instrumental differences introduce uncertainties in forecast targets. Such insights are critical for improving prediction reliability. Second, by classifying events following an approach comparable to NOAA’s S-scale\footnote{\url{https://www.swpc.noaa.gov/noaa-scales-explanation}} (S1–S5, with events in this study falling between S1 and S4), we assess how SPE properties vary systematically across different intensities. This framework not only provides an operationally meaningful benchmark but also enables testing whether discrepancies between L1 and GEO measurements depend on event magnitude. Furthermore, by exploring how event properties are influenced by geospace conditions such as local conditions at GEO and geomagnetic indices (\textit{Kp, Ap, AE, Dst,} and \textit{solar wind flow pressure}) describing the state of Earth’s magnetosphere, we aim to better understand how 10 - 50\,MeV proton precipitation into the magnetosphere is modulated. These analyses contribute to a deeper understanding of SEP transport in the coupled cislunar and near-Earth environment. 
   
    Our methodology builds on prior analyses of GOES proton flux data from SCs 22, 23, and 24, detailed in \cite{me1}, applying the same event detection methods to available SOHO/EPHIN proton flux data from SC 23 and most of SC 24 (1995 - 2016). Using NOAA S-scale thresholds, we compile a catalog of $\geq$10\,pfu SPEs at L1 in the 10 - 50\,MeV range, and generate a comparative catalog of events observed by \textit{both} instruments. All analyzed events with their S-scale classifications based on the GOES 10 - 50\,MeV proton flux range are listed in Figures \ref{sample-list} and \ref{sample-list2}. In total, we identified 40 S1 events (peak flux 10–100 pfu), 26 S2 events (100–1,000 pfu), 11 S3 events (1,000–10,000 pfu), and 6 S4 events ($>$10,000\,pfu) across SCs 23 and 24. These classifications form the basis of our analysis, allowing us to compare how SPE properties evolve with increasing intensity and how such variations affect their detection and manifestation at GEO. We then examine the statistical properties of SPEs across the S1–S4 categories and assess the potential influence of Earth’s magnetosphere on proton fluxes measured by GOES. To quantify the relationships between key parameters, we evaluate correlation strength and validity using the Kendall Tau (K$\tau$) and Spearman rank ($\rho$) correlation coefficient methods. By explicitly analyzing event properties through the lens of S-scale categories, this work highlights how SPEs of different strengths are observed at L1 and GEO, offering insights into SEP transport and the influence of Earth’s magnetosphere.
 
\section{Data preparation and products} 
    \subsection{GOES proton flux data} \label{goes_data}
    The GOES proton flux measurements are valued for their coverage of SEP events and are widely used for scientific and operational space weather applications. Our previously developed SPE-detection algorithm used $\geq$10 MeV proton flux data product from NOAA's GOES satellites (GOES-08 to GOES-15) spanning SCs 22-24 \citep{me1}. Positioned $\sim$36,000 km above Earth's equator, GOES provides almost uninterrupted proton flux measurements crucial for space weather forecasting \citep{pl_example_aminalragia}. To ensure data consistency, NOAA designates a ``primary'' satellite during the periods when several spacecraft observe. Similarly, we select a ``primary'' instrument for each month based on an empirical approach, giving priority to the one measuring the higher peak proton flux during the SPEs of that month. Other studies like those by \citet{pl_example_rotti} and \citet{pl_example_aminalragia} apply different criteria to select ``primary'' and ``secondary'' satellites. With the launch of GOES-13 in 2011, the Energetic Proton, Electron, and Alpha Detector (EPEAD) replaced the Energetic Particle Sensor (EPS), providing similar integrated SEP data products. Unlike the EPS, which was single-direction-oriented, the updated EPEAD now features detectors facing both East and West. For this study, we compare the fluxes recorded by each detector and utilize proton flux data from the primary instrument and a single detector on it (either East or West), whichever reports higher proton fluxes. This approach avoids the simple averaging of fluxes that could misrepresent SEP characteristics at GEO. Traditionally, GOES integrated proton flux products over the following energy ranges are considered for the analysis: $\geq$1 MeV, $\geq$5 MeV, $\geq$10 MeV, $\geq$30 MeV, $\geq$50 MeV, $\geq$60 MeV, and $\geq$100 MeV. Correspondingly, the fluxes recorded for the $\geq$1 MeV data product include contributions from all higher energy channels, including those above 100 MeV. While most of the GOES data are corrected for galactic cosmic background and high-energy particles penetrating the shielding, NOAA warns about potential cross-contamination in the proton detectors during SPEs.
    
    
    \subsection{EPHIN proton flux data}
    EPHIN, a component of the COmprehensive SupraThermal and Energetic Particle Analyser \citep[COSTEP,][]{costep_muller} aboard SOHO at L1, was designed to enhance the understanding of energy deposition and particle acceleration in the solar environment \citep[][]{ephin_use1_laurenza,ephin_noise_kuhl,review_whitman}. EPHIN produces approximately 15 megabytes of scientific data daily, which can be used to study the energy spectra of electrons, protons, and helium nuclei. Charged particles are detected through ionization within the sensor, which utilizes a multi-element array of solid-state detectors with anticoincidence, which enables accurate energy measurements. For high-energy proton detection, EPHIN uses differential energy channels spanning 3.98 to 50.12 MeV \citep{costep_muller}. Differential proton flux data from EPHIN have advanced the study of solar particle acceleration and transport processes, as well as in monitoring space weather phenomena that can affect satellite operations and communications on Earth \citep[e.g.,][]{ephin_eg1_kuhl,ephin_eg2_posner,ephin_eg3_gomez}. It is especially important that EPHIN is located beyond Earth's magnetosphere and is representative of cislunar space, enabling the monitoring of SEP dynamics before they interact with the magnetosphere. In this work, we are utilizing the proton flux measurement products from SOHO/EPHIN that were obtained using the Relativistic Electron Alert System for Exploration (REleASE) algorithm developed by \citet{goes_contamination_posner}, and accessed through the dataset published by \citet{zenodo}\footnote{\url{https://zenodo.org/records/14191918}}. This dataset, generated from Level 1 EPHIN data collected between 1995 and 2016, includes proton fluxes across 22 energy bins and forms the basis of the EPHIN measurements used in our analysis.  
  
    EPHIN measurements can be impacted during strong SPEs, albeit to a lesser extent. According to the documentation of EPHIN Level 3 data products \citep{ephin_doc}, the instrument is designed with an inner and outer ring at the detector site to capture and count incoming protons. However, during intense SPEs, the detector can become overwhelmed by the high volume of incoming particles, causing the outer ring to shut off. This results in an abrupt reduction in the effective geometric factor, which corresponds to an immediate drop in recorded counts. Because many flux products are normalized by geometric factors or rely on continuous count accumulation, such sudden changes in the detection configuration can manifest as step-like discontinuities (jumps) in derived quantities like particle flux (i.e., count rate per geometric factor, as described in \citep{costep_muller}). Currently, no established method exists to fully mitigate these instrumental effects. However, these flux variations are generally $\lesssim$50 pfu and do not significantly impact our analysis when comparing SPE characteristics between L1 and GEO. During our visual inspection of the SEP events shared below, we identified only one instance where a falsely elevated flux surpasses the true peak flux of an event, which was corrected in the catalog.

    The two instruments, GOES and EPHIN, measure proton fluxes at different differential channels and produce distinct observational data products, necessitating adjustments to make their data comparable. It is important to note that EPHIN data products routinely report protons with energies up to only $\sim$50 MeV. As a result, to ensure consistency with EPHIN's measurable energy spectrum, we process GOES integral proton fluxes to exclude fluxes above 50 MeV, limiting the flux range to 10 - 50\,MeV. This was done by subtracting the integral flux product of $\geq$50 MeV from $\geq$10 MeV. GOES differential channel measurements are available at 1-minute cadence; however, we utilize the integral data products ( $\geq$10 MeV,  $\geq$50 MeV,  $\geq$60 MeV, $\geq$100 MeV), which are produced by a NOAA internal algorithm that accumulates data over 5-minute intervals. For consistency, EPHIN data were averaged to 5-minute intervals to align with the GOES measurements. With the 10 – 50 MeV fluxes now isolated for both instruments, the datasets are suitable for direct comparative analysis.
    
    \subsection{A catalog of concurrently detected SPEs at GEO and L1} 
    \begin{figure} 
\centering 
  \includegraphics[width=.99\textwidth]{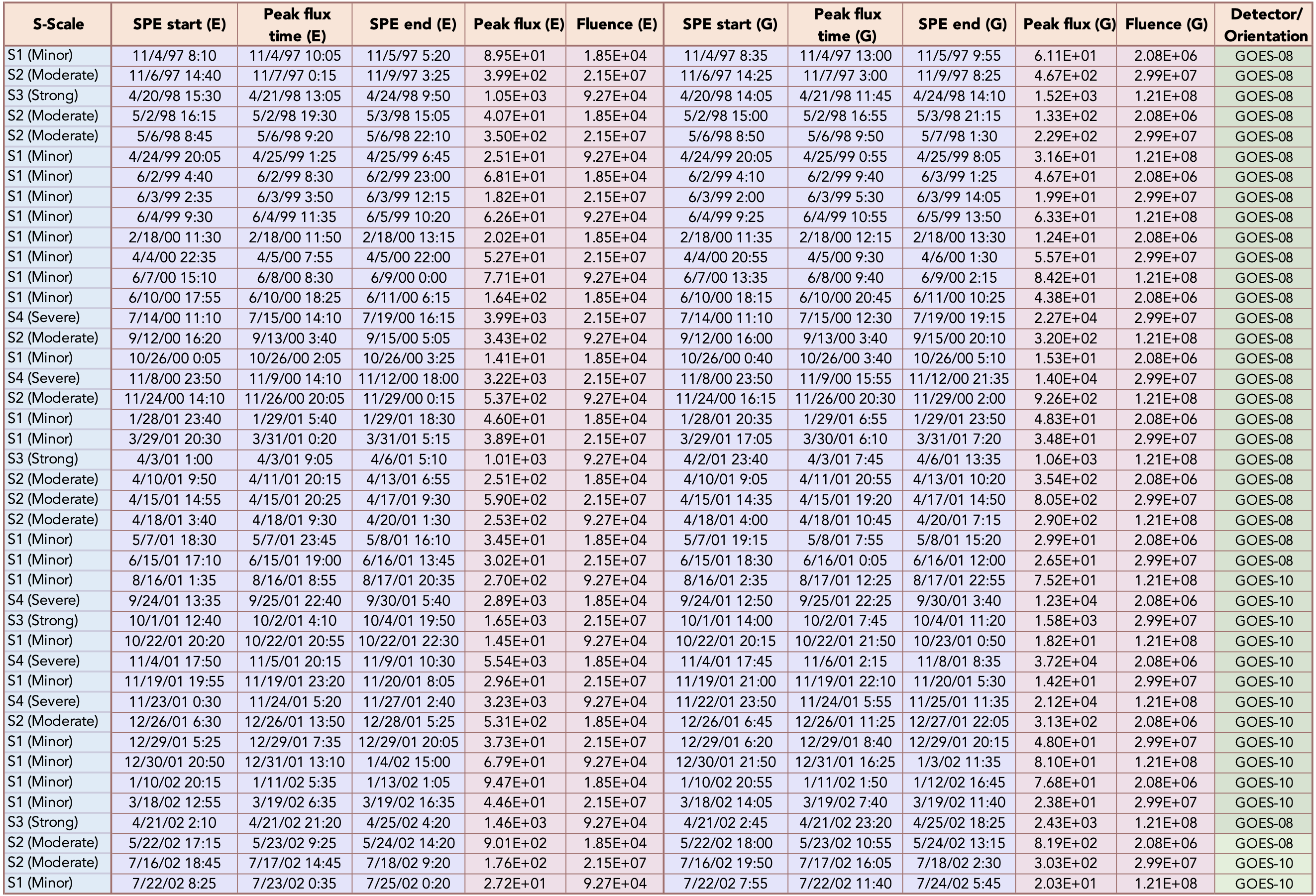} 
  \caption{List of 83 SPEs observed concurrently by EPHIN and GOES during SC 23 and the majority of SC 24. The first column provides the NOAA-like S-scale classification for each event based on its peak flux, while the last column specifies the GOES instrument designated as the primary detector at the time of observation. Reported peak fluxes are given in pfu, and fluences in pfu$\cdot$s.} \label{sample-list}
\end{figure}

\begin{figure} [!ht]
\centering 
  \includegraphics[width=.99\textwidth]{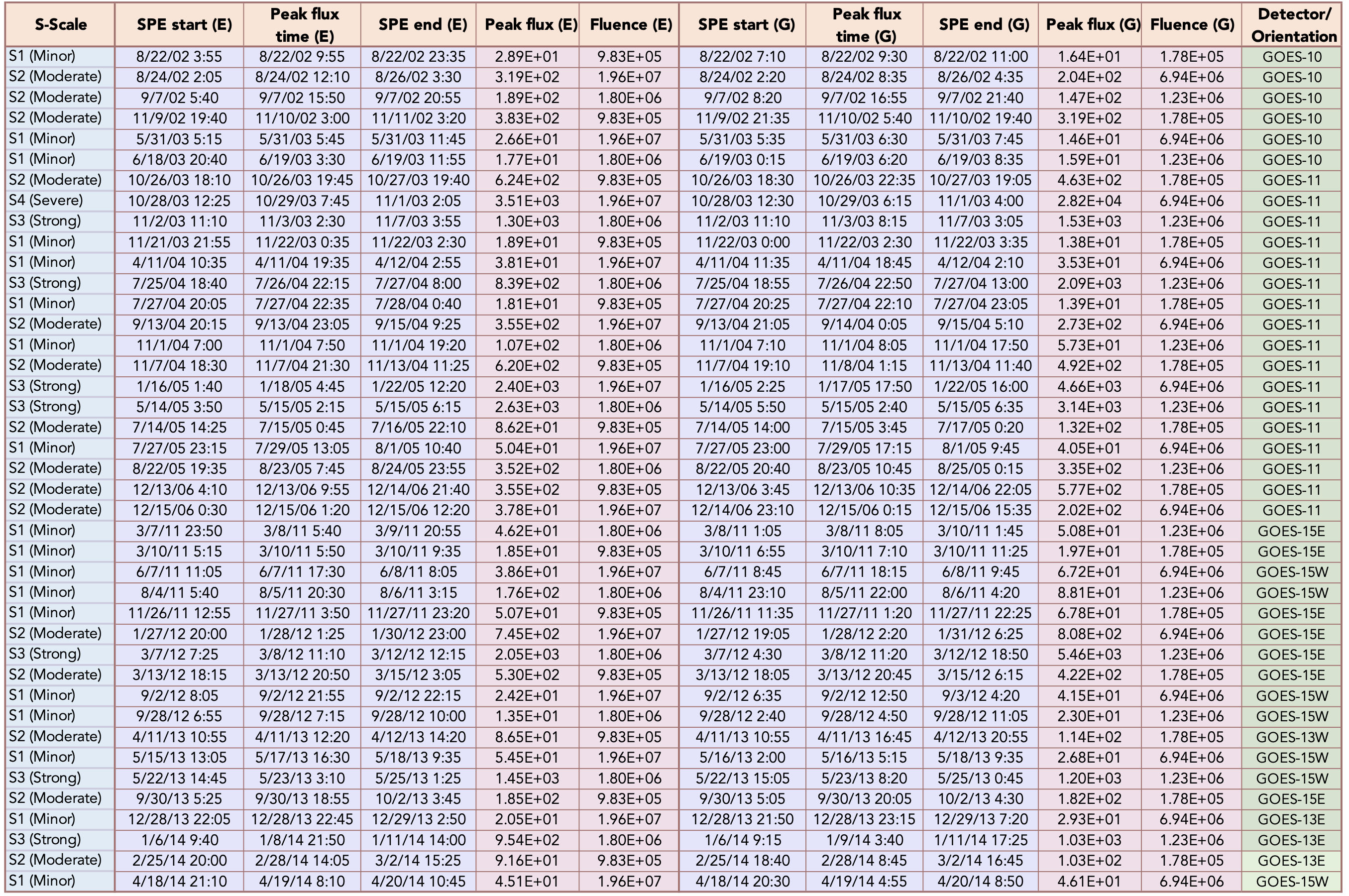} 
  \caption{Fig. \ref{sample-list} continued.} \label{sample-list2}
\end{figure}    

    Analogously to NOAA's definition of an SPE, we record SPEs as instances where at least 10 pfu of protons with energies 10 - 50\,MeV are detected for at least 15 minutes (correspondent to three subsequent data points). The events are recognized separately for GOES and EPHIN. We document SPE properties such as start and end times, peak flux and its timing, and event fluence as measured by each instrument. To prevent overcounting events due to flux oscillations near the 10 pfu threshold, closely spaced events ($\leq$10 minutes apart) were merged, as these likely reflect minor variations in a single event rather than multiple separate events. When merging events, relevant parameters (e.g., start and end times) were adjusted accordingly. The finalized catalog of the SPE events is presented in Figures~\ref{sample-list}~and~\ref{sample-list2}. 
\begin{figure} [!htp]
\centering
 \includegraphics[width=.85\textwidth]{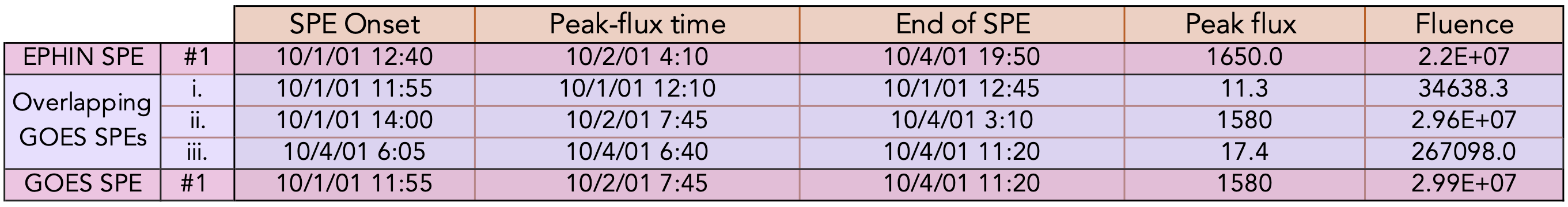}
  \caption{Merging 3 GOES-detected SPEs into one as they \textit{all} occur during the same (singular) event detected by EPHIN.}
    \label{tab:overlap}
\end{figure} 

\begin{figure}
\centering 
 \includegraphics[width=.75\textwidth]{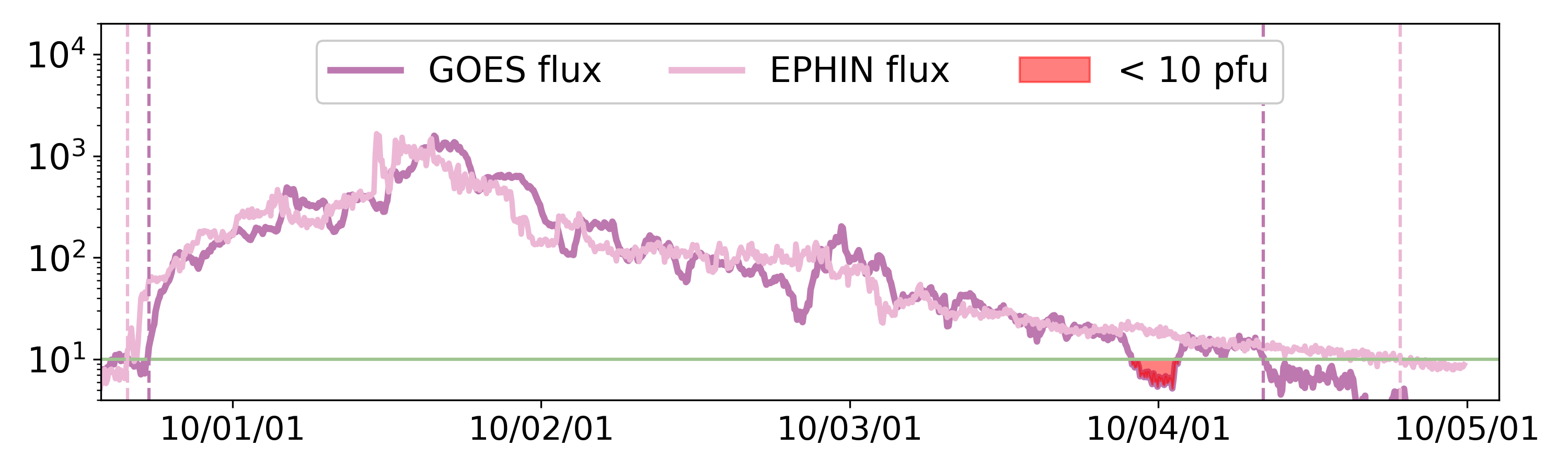}
  \caption{Flux profile of the SPE shown in Fig. \ref{tab:overlap}, after applying the adjustments described therein. The vertical lines mark the event intervals identified by each instrument, while the red-shaded regions highlight periods when GOES fluxes momentarily drop below 10 pfu but rise back above this threshold within 10 minutes.}
  \label{fig:writeup_event}
\end{figure}

    We note that isolated events- events detected by either EPHIN \textit{or} GOES but not both- often only marginally exceed the 10 pfu threshold. In this study, we focus exclusively on SPEs observed concurrently by both instruments, excluding cases detected solely by GOES or EPHIN. Consecutive detections of multiple SPEs by one instrument were treated as a single event if identified as such by the other instrument (i.e., overlapping with a single event from the other instrument). Figure \ref{tab:overlap} presents an example of SPE records detected by GOES that corresponded to a single SPE event in EPHIN and therefore merged. Figure \ref{fig:writeup_event} demonstrates the SPE detection method itself, highlighting the interval where GOES fluxes shortly drop below 10 pfu.
    
    \section{Comparing properties of SPEs detected by GOES and EPHIN}  \label{comparing}  
    \begin{figure} 
\centering 
\includegraphics[width=.95\textwidth]{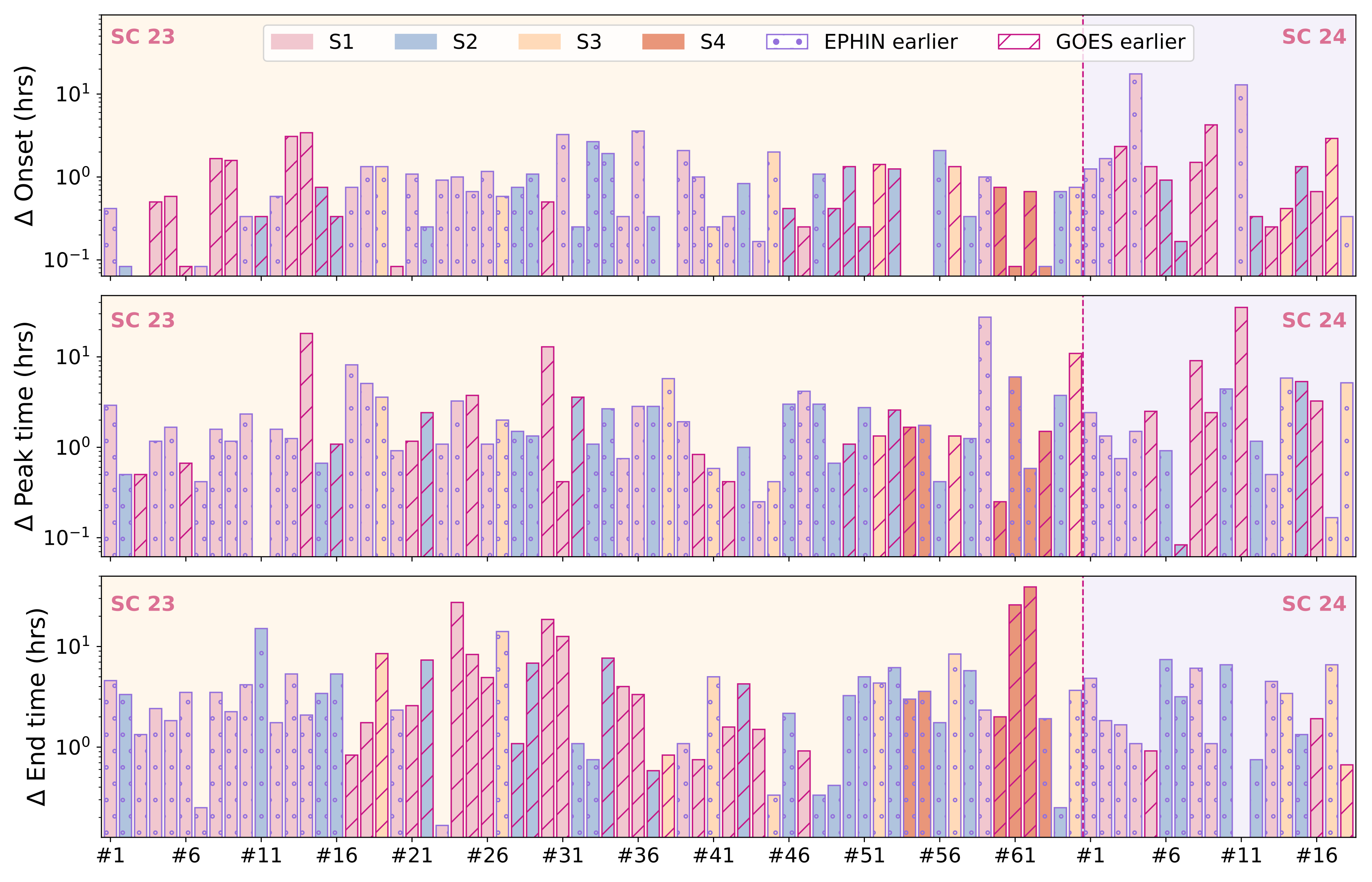} 
   \caption{Timing offsets in SPE onset (top), peak flux (middle), and end times (bottom) as observed by GOES and EPHIN during SCs 23 \& 24. Each bar corresponds to a single event, with the x-axis enumerating events in chronological order. Bar colors denote the NOAA S-scale intensity, while hatching indicates which instrument recorded the earlier time. Shaded backgrounds distinguish SC 23 and SC 24. Instances where bars are absent (e.g., onset for event \#3) indicate no measurable offset, i.e., both instruments detected the property simultaneously ($\Delta = 0$ hrs).}
 \label{deltas}
\end{figure}
     
    With our SPE sample split into the 4 S-scales based on the event peak fluxes in 10 - 50\,MeV range according to GOES measurements, we compare 10 - 50\,MeV SPE flux-related properties measured by the EPHIN and GOES instruments. Due to their often-presented large dynamic range, for the quantities that can have positive or negative times, we apply a signed logarithmic transformation to the differences between their measurements. The transformation is defined as:
    
     \begin{equation}
    \mathrm{SignedLog}(\textbf{x}) = \mathrm{sign}(\textbf{x}) \cdot \log_{10}(|\textbf{x}| + \lambda).
    \end{equation}
    
    Here, \textbf{x} represents the difference between EPHIN and GOES measurements for a given SPE parameter (e.g., start time in minutes). A small positive constant, \textbf{$\lambda$} (set to 1), is added to avoid undefined logarithmic values near zero, while $\mathrm{sign}(\textbf{x})$ preserves direction: positive values indicate \textit{GOES records earlier times}, negative values indicate \textit{EPHIN records earlier times}. Because the logarithmic scale accommodates wide dynamic ranges, it is especially useful for comparing peak flux and fluence, where instrument discrepancies can be substantial. \begin{figure}[ht!]
\centering 
\includegraphics[width=.99\textwidth]{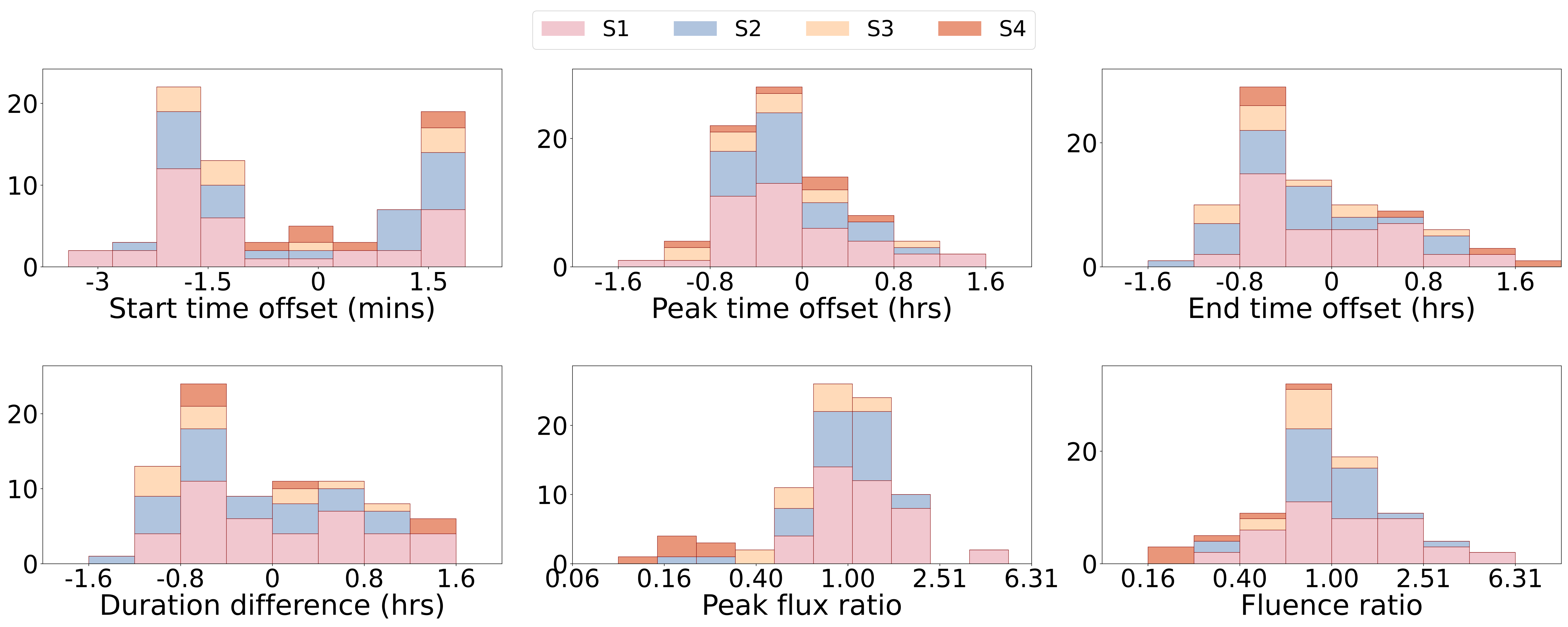} 
   \caption{Signed logarithmic differences in SPE properties detected by EPHIN and GOES. Shown are differences in SPE (a) start time, (b) peak flux time, (c) end time, (d) duration, (e) peak flux ratio, and (f) fluence ratio.} \label{fig:hists}
\end{figure}
\begin{figure}[ht!]
\centering 
\includegraphics[width=.99\textwidth]{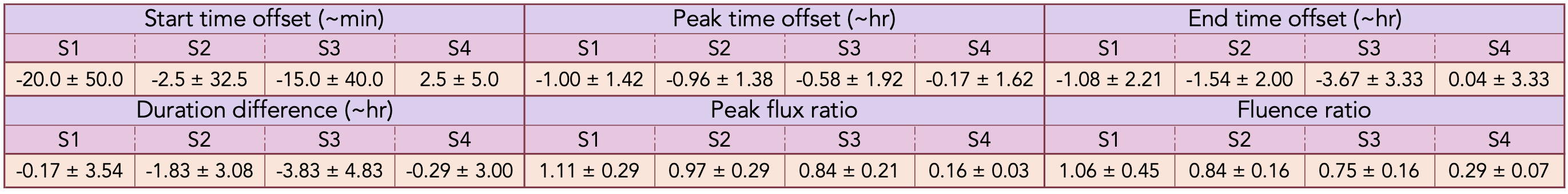} 
   \caption{Signed logarithmic differences in SPE properties between EPHIN and GOES by S-scale. Negative values indicate earlier onsets, peaks, or ends (or shorter durations) for EPHIN, while ratios greater than one denote higher fluxes measured by EPHIN.} \label{fig:offset_num}
\end{figure}

    Differences in event start times, peak flux times, and end times are summarized in Figure~\ref{deltas}. The differences are represented using the median values for individual events, with the median absolute deviations (MADs) providing the measure of variability. We found that the mean values for the quantities significantly differ from the medians, revealing potentially-skewed distributions; therefore, the median values serve a better quantification of differences for the current data set. SPE onsets show that EPHIN generally detects S1--S3 earlier than GOES, with per-scale medians (EPHIN relative to GOES) of $-20\pm50$\,min (S1), $-2.5\pm32.5$\,min (S2), and $-15\pm40$\,min (S3), consistent with its vantage point outside Earth’s magnetosphere. By contrast, S4 onsets show a small positive median of $+2.5\pm5$\,min, indicating slightly earlier detections by GOES. Notably, the S1 onset offset ($\sim20$\,min median) is far longer than the transit time of 10\,MeV protons from L1 to GEO and is instead comparable to solar-wind propagation times. Similar trends appear for peaks, ends, and overall durations. EPHIN recorded earlier peaks for S1--S3 with medians of $-1.00\pm1.42$\,hr, $-0.96\pm1.38$\,hr, and $-0.58\pm1.92$\,hr, respectively, while S4 peaks were near-simultaneous ($-0.17\pm1.62$\,hr). Event ends follow the same pattern, with EPHIN earlier for S1--S3 ($-1.08\pm2.21$\,hr, $-1.54\pm2.00$\,hr, and $-3.67\pm3.33$\,hr) and GOES slightly earlier for S4 ($+0.04\pm3.33$\,hr). Durations were longer at L1 for S1--S3 (medians $\sim-0.17$ to $-3.83$\,hr) and comparable for S4 ($-0.29\pm3.00$\,hr). Flux comparisons show higher particle counts by EPHIN for S1 (peaks $\sim1.12\pm0.30$, fluence $\sim1.06\pm0.45$), while ratios are suppressed at higher intensities: peaks $\sim0.97\pm0.29$, $0.84\pm0.21$, $0.16\pm0.03$ and fluence $\sim0.84\pm0.16$, $0.75\pm0.17$, $0.29\pm0.07$ for S2, S3, and S4, respectively. Distributions are shown in Figure~\ref{fig:hists}, and Figure~\ref{fig:offset_num} presents the full set of medians and MADs on a signed logarithmic scale.  

    Overall, EPHIN shows a systematic tendency to register earlier onsets, peaks, and ends for S1--S3, while this timing advantage diminishes for S4, where both instruments detect nearly the same number of cases. In terms of particle counts, S1 events favor EPHIN, with peak flux and fluence median ratios of 1.12$\pm$0.30 and 1.06$\pm$0.45, respectively. These ratios may point to magnetospheric deflection effects: among the 40 S1 events, 22 exhibited peak flux ratios and 21 exhibited fluence ratios above 1 (favoring EPHIN), whereas 18 showed peak flux ratios and 19 showed fluence ratios below 1 (favoring GOES). The S3--S4 event categories result in larger GOES fluxes, and S2 is split as peak fluxes slightly favor GOES ($0.97\pm0.29$) whereas fluences favor EPHIN ($0.84\pm0.16$).

    As studied by \citet{goes_contamination_posner}, contamination of GOES fluxes by high-energy protons and electrons may account for some of the counterintuitive results in Figure \ref{fig:hists}, where GOES occasionally recorded key SPE timing properties earlier than EPHIN and, at times, substantially higher proton counts. High-energy particles, such as $>$100 MeV protons, travel faster and reach GOES detectors earlier than protons in the 10 - 50\,MeV range, which we focus on in this analysis. When these higher-energy particles contaminate lower-energy channels, the $\geq$10 MeV threshold for SPE detection can be crossed prematurely, causing events to be flagged earlier than expected. This effect can also affect the timing of peak fluxes, the magnitude of measured peak fluxes, and the integrated event fluence. We observe substantial differences in peak flux and fluence values, with GOES often reporting higher levels during stronger events (S4 storms). Such discrepancies may arise from localized enhancements, preferential particle transport, or contamination effects in the flux measurements. Overall, although 10\,MeV protons can theoretically cover the distance from L1 to Earth in about a minute (L1 and GEO being $\sim$1.5 million km apart), we see that delays in some SPE onset times can extend to several hours. Delays in peak flux times are also typically longer than those observed for onset times, with several events showing delays exceeding 10 hours. The time discrepancies can be considered intrinsic uncertainties in determining key SEP timing properties from a given observational vantage point or instrument (unless given a physical explanation). Consequently, SPE prediction efforts for $\geq$10 MeV event onset and peak times are likely subject to an inherent precision limit of around 30 minutes due to these observational differences, potentially limiting short-term forecasting efforts for lower-energy SEPs.
    
    Differences in event onsets and durations based on event strength as observed by EPHIN may reflect its vantage point at L1, offering stronger magnetic connectivity to SEP acceleration regions, whereas GOES at GEO is subject to Earth’s magnetospheric shielding, which can inhibit particle access and prolong observed event decay. \citet{bruno2018} found that high-energy ($>$80 MeV) SEPs from poorly connected events are typically long ($>$6 days) and weak, with extended durations attributed to cross-field diffusion and co-rotation with the Sun. Similarly, \citet{connectivity2024} observed that spacecraft with stronger magnetic connectivity to SEP event source regions measured higher proton fluxes and shorter event durations, while those with weaker connectivity observed lower fluxes and longer durations. They suggest that spacecraft with poor magnetic connectivity encounter more complex magnetic field configurations, which delay particle arrival and extend event durations. In contrast, spacecraft with better connectivity experience more direct particle propagation along well-connected magnetic field lines, resulting in shorter durations.
  
    \begin{figure}[ht!]
\centering 
 \includegraphics[width=1.0\textwidth]{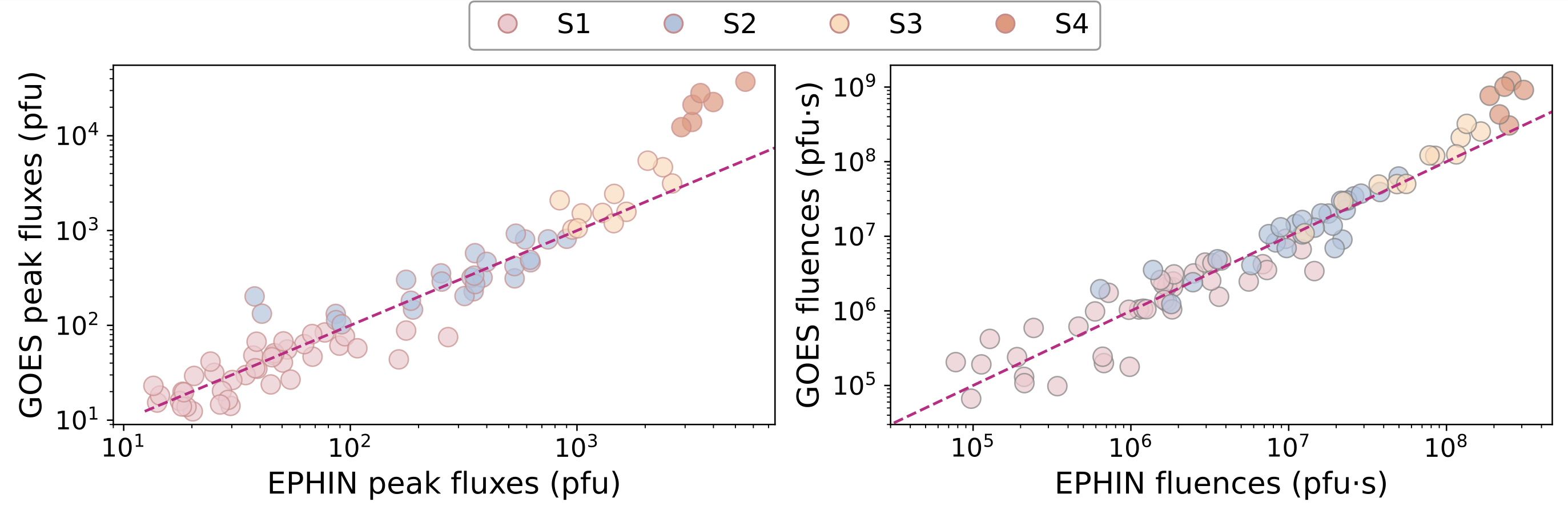}
    \caption{Comparison of GOES peak fluxes (left) and fluences (right) with EPHIN during SPEs across SCs 23 \& 24.}
  \label{fig:ratio}
\end{figure}


    Figure \ref{fig:ratio} presents a comparison of peak fluxes and fluences between EPHIN and GOES proton measurements. The dashed magenta line marks the one-to-one reference. Points near this line indicate good agreement between instruments, while large deviations highlight events where peak fluxes differ substantially, reflecting instrumental or environmental effects. In the left panel, weak events (S1–S2) cluster near the one-to-one line, generally indicating good agreement between EPHIN and GOES peak fluxes at lower intensities. S3 events also remain close to unity, mostly reflecting consistency between L1 and GEO for moderately strong events (although visibly demonstrating statistically higher GOES peak fluxes). By contrast, S4 events lie well above the line in the upper-right, where GOES records substantially higher peak fluxes than EPHIN. This supports the possibility that flux contamination amplifies the 10 – 50 MeV proton signal by up to an order of magnitude during the strongest SPEs; discussed later in Section~\ref{issues}. A similar pattern appears in integrated fluence; while most events follow the one-to-one trend, several lie well above it, with GOES fluence exceeding EPHIN by as much as an order of magnitude. This systematic overestimation shows that contamination may affect not only peak fluxes but also cumulative fluence, underscoring how GOES proton channels can misrepresent SPE intensities- key parameters for operational forecasting. Identifying and correcting such cases is therefore critical when interpreting near-Earth proton data. Discrepancies in SPE properties also likely stem from a range of complex, dynamic processes, including magnetic connectivity, particle transport efficiency, the influence of interplanetary magnetic structures, drift motion, diffusion, and turbulence \citep{transport_effects_battarbee}.
    
    Overall, the results presented in this section highlight the importance of accounting for both instrument- and location-specific effects when analyzing SEP data. Differences in onset times, flux levels, and event durations can complicate the extrapolation of observations from one location to another, potentially degrading the accuracy of space weather forecasts. Incorporating these considerations into predictive models would substantially improve forecast reliability, particularly for near-Earth and lunar mission operations.

    \section{Analyzing effects of the Geospatial environment} \label{magnetosphere}

    Variations in SEP properties such as speed, density, and energy naturally contribute to differences in measured proton fluxes, alongside factors like instrument position, orientation, sensor sensitivity, and directionality. A significant consideration in our analysis is the potential influence of Earth's magnetosphere and its dynamics, which may introduce magnetic shielding effects at GEO, including particle scattering and reduced flux detection during SPEs; effects that are absent in the cislunar region. For example, \citet{magnetospheric_shielding_kress} explored the impact of solar wind dynamic pressure and interplanetary magnetic fields (IMFs) on SEP precipitation into GEO, and reported rapid SPE onsets observed by the GOES West EPEAD compared to simultaneous flux measurements from the East detector, attributing this discrepancy to differences in geomagnetic shielding. Their findings show significant shielding effects for SEPs in the energy ranges of 6.5 – 12.5 MeV for the West detector and 40 – 80 MeV for the East detector. Further, \citet{kress_cutoff} demonstrated that during SPEs, rigidity cutoffs (defined as the minimum momentum per unit charge required for particles to precipitate to the GEO orbit) can be suppressed, allowing lower-energy particles to be detected by GOES. In addition, \citet{particle_energy_obrien} emphasized the central role of particle energy in determining SEP access across different regions of Earth’s magnetosphere. Together, these studies underscore how multiple aspects of the magnetosphere can substantially shape SEP propagation and detection at GEO.
    
    To explore whether geospatial properties are related to the EPHIN-to-GOES peak flux, we employ $K\tau$ and Spearman $\rho$ methods. $K\tau$ and Spearman $\rho$ assess monotonic relationships without assuming a specific functional form between variables. Both capture whether changes in one variable consistently correspond to changes in the other, regardless of the relationship’s shape or strength. Each method also yields a p-value, which reflects the probability of incorrectly rejecting the null hypothesis of no correlation with respect to the hypothesis of the presence of correlation. A p-value $<$ 0.05 is typically referred to as a sign of a statistically significant relationship.

    \citet{arrival_direction_filwett} emphasizes the auroral electrojet (\textit{AE}) index as a key driver of magnetospheric cutoff boundaries, shaping the particle populations that can access GEO. They further demonstrate that proton penetration into the near-equatorial inner magnetosphere is highly sensitive to storm-time IMF conditions and solar wind drivers, which dynamically alter magnetospheric transmission properties. More recently, \citet{magnetospheric_shielding_kress} utilized NOAA’s advanced Solar Proton Sensor to track real-time cutoff variations, confirming that elevated geomagnetic activity directly influences SEP access at GEO. Together, these findings highlight correlations where geomagnetic indices like \textit{AE, Kp, Ap, Dst} and dynamic cutoff shifts govern SEP penetration, particularly during contaminated SPEs. To investigate similar relationships, we analyze publicly available OMNI2 data\footnote{\url{https://spdf.gsfc.nasa.gov/pub/data/omni/low_res_omni/}}, which provides hourly averaged geomagnetic indices from spacecraft near Earth’s orbit. Specifically, we consider the following indices: 

    \begin{itemize}
      \item (AE) Index: Reflects magnetic activity in the auroral zone produced by enhanced ionospheric currents;
      \item Planetary Geomagnetic Activity (\textit{Kp}): A 0–9 scale quantifying global geomagnetic activity every three hours based on the oscillations of the horizonal component of the geomagnetic field;
      \item Planetary Amplitude (\textit{Ap}): The daily average linear measure of geomagnetic activity derived from \textit{Kp};
      \item Disturbance Storm Time (\textit{Dst}): The hourly change in Earth’s low‐latitude horizontal magnetic field, indicating ring‐current strength during storms;
      \item Flow pressure: The solar wind dynamic pressure at the L1 point, calculated as:
        \[P \;=\; \frac{1.67}{10^{6}}\;N_{p}\,V^{2}\,\bigl(1 + 4\,\tfrac{N_{\alpha}}{N_{p}}\bigr),\]
        combining proton density \(N_{p}\), bulk solar-wind speed \(V\), and the alpha-to-proton ratio \(N_{\alpha}/N_{p}\) (the fraction of helium ions relative to protons).
    \end{itemize}
 
\begin{figure}[ht!]
\centering 
 \includegraphics[width=.95\textwidth]{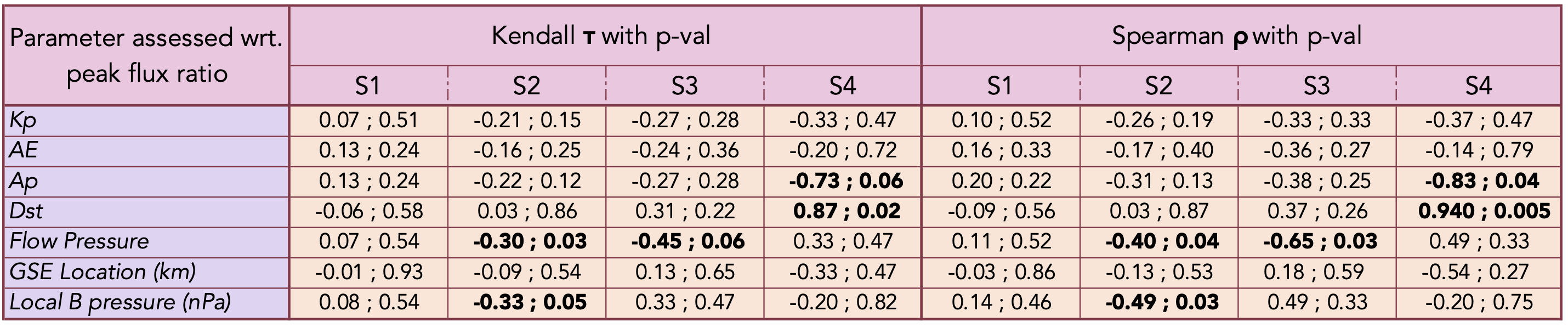}
  \caption{Correlation analyses of SPEs with geomagnetic indexes, GOES positioning, and local magnetic pressure at GEO.}
  \label{geomag}
\end{figure}

    To evaluate whether these parameters generally modulate the particle precipitation from L1 and GEO, we examine them in relation to the EPHIN-to-GOES peak flux ratio. Each geomagnetic index is evaluated using its event-averaged value. We report K$\tau$ and Spearman $\rho$ values along with their corresponding p-values in Figure \ref{geomag}. 

    
    The analyses reveal very few statistically significant correlations (p $\leq$ 0.05, shown in bold) between flux ratios and geomagnetic indices, and even fewer cases where these correlations are both statistically significant \textit{and} of meaningful strength (coefficients $<$–0.5 or $>$0.5). Notably, S4 events are the only subset to exhibit such relationships: correlations with \textit{Ap} reach –0.73 (p = 0.06) and –0.83 (p = 0.04) for Kendall $\tau$ and Spearman $\rho$, respectively, while correlations with \textit{Dst} reach 0.83 (p = 0.02) and 0.94 (p = 0.005). The consistency across methods suggests a possible link between flux measurements and geomagnetic activity for $\geq$ S2 events. However, while the p-values indicate significance for some geomagnetic properties, the coefficients themselves reflect only moderate associations, implying that the influence of geomagnetic conditions on these variations is limited and should be interpreted with caution. If \textit{Ap} were strongly associated with the flux ratios, a comparable correlation with \textit{Kp} would also be expected; however, no such pattern emerges. Likewise, although \textit{Dst} yields relatively strong correlation coefficients, these signals are sporadic and not consistently mirrored in other geomagnetic indices, calling into question the robustness of these relationships. Overall, we suggest interpreting with caution even those correlations in this work that are statistically significant based on the p-value.

    Moreover, SEP events are fundamentally driven by solar processes such as coronal mass ejections (CMEs) and solar flares and should not have a direct link to the geomagnetic activity besides the transport effects. Geomagnetic indices like \textit{Ap}, \textit{Kp}, or \textit{Dst} reflect the magnetospheric response to interplanetary disturbances such as CMEs rather than serving as causes of SEP generation. For instance, \citet{geomag1} demonstrated correlations between eruptive magnetic flux from CMEs and geomagnetic indices such as \textit{Ap} and \textit{Dst}, but characterized these indices as responses of Earth’s magnetosphere to solar eruptions, with no direct role in SEP production. Similarly, \citet{geomag2} analyzed geomagnetic field variations during SEP events in SC 23 and found correlations with high-latitude indices (\textit{AE} and the Polar Cap index) during major events, indicating geomagnetic responses to SEPs and associated CMEs rather than predictive or causal drivers. Thus, even when statistically significant correlations emerge between flux ratios and geomagnetic indices (as in our case with \textit{Ap} and \textit{Dst}), their interpretation remains ambiguous and might not indicate any causality. The absence of consistent correlation with the remaining geomagnetic measures suggests that these relationships may be incidental rather than a direct or reliable modulator of SEP characteristics. Taken together, the presence of statistically detectable but moderate correlations suggests that such relationships are unlikely to provide strong predictive value (in terms of their precipitation to the geostationary orbit) for SPE forecasting \citep{coeffs}.  

    Next, we explore whether the location of the GOES satellite inside the magnetosphere during SCs 23 \& 24 coincided with increased detections of weaker or stronger peak flux ratios. The basic idea is that the magnetic field environments for GOES differ at the dayside and nightside of the magnetosphere. Using GOES ephemeris data in Geocentric Solar Ecliptic (GSE) coordinates, we determined the spacecraft’s position for all 83 SPEs. Comparing the EPHIN-to-GOES peak flux ratios with the satellite’s location revealed no clear correlation between event properties and specific orbital positions on either the ``dayside'' or ``nightside'' positioning. Lastly, we examine whether the ratio of proton fluxes correlates with variations in the local magnetic pressure at GEO. For this, we utilize data from the GOES magnetometer\footnote{\url{https://www.swpc.noaa.gov/products/goes-magnetometer}}, which provides vector components of the magnetic field: (i) ${H}{p}$ pointing northward and parallel to Earth’s spin axis, (ii) ${H}{e}$ perpendicular to ${H}{p}$ and ${H}{n}$, directed earthward, (iii) ${H}{n}$ perpendicular to both ${H}{p}$ and ${H}_{e}$, pointing eastward, and (iv) the total field strength. These magnetometer data have been widely used for purposes ranging from analyzing energy release in Earth’s magnetosphere during geomagnetic storms to monitoring solar wind conditions during extreme solar storms when GOES crosses the magnetopause. However, no significant correlation was found between local magnetic pressure and the EPHIN-to-GOES flux ratio.

    In addition to the analysis of the peak flux ratios, we have also performed the analysis of the fluence ratios for the detected SEP events, as well as checked how the utilization of the properties at the time of the GOES peak flux compares to the event-averaged properties. The statistically significant correlations for this analysis, if present, were still of a sporadic nature, and therefore, we do not provide the details of these correlations here.

\section{Discussion on the potential contamination in GOES flux} \label{issues}  
\begin{figure}[ht!]
\centering 
 \includegraphics[width=.85\textwidth]{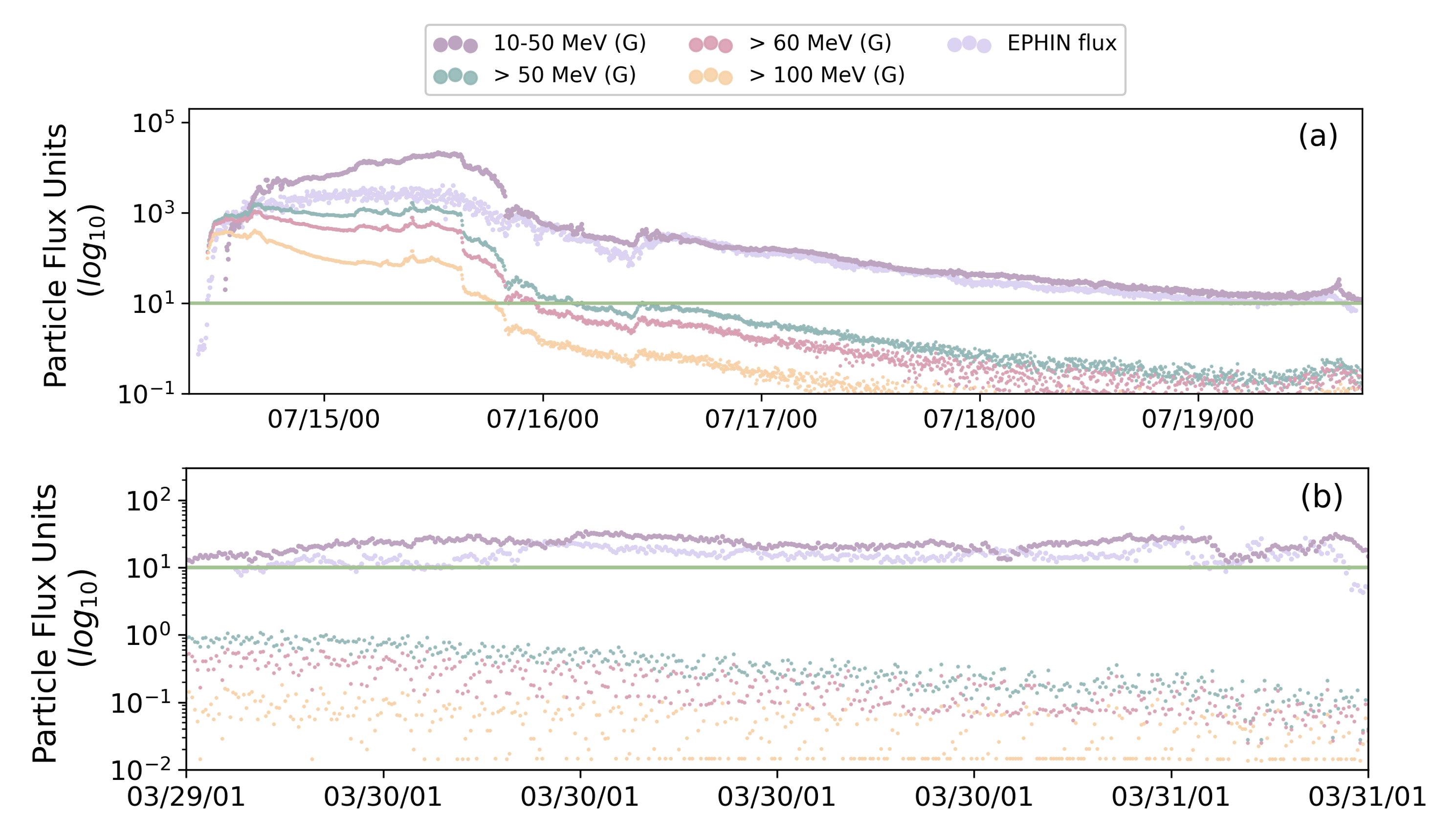}
  \caption{(a) An SPE where the 10 – 50 MeV proton flux profile closely mirrors the signals in the higher-energy GOES channels, indicating potential contamination. (b) An SPE showing no obvious contamination signatures, with proton channels above 50 MeV showing significantly lower signals compared to the 10 – 50 MeV channel, as expected.}
  \label{fig:issue-and-fix}
\end{figure}

During intense SEP events, lower-energy channels of GOES instruments can sometimes record falsely elevated signals due to contamination from high-energy particles, such as relativistic electrons or protons in the upper MeV to GeV range \citep{goes_contamination_posner}. To examine the possible imprint of flux contamination effects, we use the following criteria to identify SPEs that may have contaminated GOES flux data:
    \begin{itemize} 
    \item If increases in the 10 - 50\,MeV fluxes occur simultaneously with those in the higher-energy channels ($>$50 MeV, $>$60 MeV, $>$100 MeV) and if there is no CME shock arrival at L1 at around that time, this raises contamination concerns. Since 10 - 50\,MeV protons should take longer to reach the detector, such early increases suggest contamination rather than realistic physical changes. 
    \item Rapid, unique changes in the flux of higher-energy channels should not be mirrored in the lower-energy channels unless contamination is present.
    \item If GOES fluxes are comparable to or larger than EPHIN fluxes, it suggests a high likelihood of data contamination. Since EPHIN is not subject to the same contamination issues as GOES, any significant or prolonged difference between GOES and EPHIN measurements during such episodes should be weak or negative. Large discrepancies favoring GOES indicate that its data may be affected. 
    \end{itemize}

We note here that these effects are based on the integral proton flux measurements and do not provide an exclusive set of contaminated and non-contaminated events. A detailed look into the differential fluxes and related timing, as well as establishing the proper quantitative criteria for contamination presence, are necessary to determine the contamination more accurately. However, we believe that this qualitative look is sufficient for the current work to estimate the contamination impact on the analysis.

Using these criteria, we visually examined the 83 SPEs in our sample and identified 17 with potential contamination, though they were still included in the analysis (listed below). Figure \ref{fig:issue-and-fix} compares these scenarios: (top) contamination of the 10 - 50\,MeV channel by higher-energy protons, and (bottom) an event without this contamination. In the top panel, the 10 - 50\,MeV measurements exceed the 10 pfu threshold only 10 minutes after the $\geq$50 MeV, $\geq$60 MeV, and $\geq$100 MeV channels surpass the same threshold. This short delay suggests that the 10 - 50\,MeV measurements may be affected by contamination from the higher-energy channels, as a longer delay would be expected if the lower-energy measurements were unaffected. Additionally, multiple abrupt changes in the higher-energy flux profiles are mirrored in the 10 - 50\,MeV profile, further supporting the idea that stronger fluxes influence the 10 - 50\,MeV measurements. The significant and prolonged difference in fluxes between EPHIN and GOES during the first day of the SPE is also unexpected and raises concerns, as it is not entirely realistic given the positioning of both instruments. Taken together, these factors lead us to classify the SPE in the top panel as potentially contaminated. In contrast, the bottom panel does not exhibit these issues. While these questionable fluxes may reflect actual SEP behavior, they also raise the possibility of additional undetected contamination in GOES data. These challenges in reliably utilizing GOES proton flux data highlight the importance of evaluating and accounting for SEP contamination, particularly during intense SEP events. In future work, we aim to develop methods to correct the affected data and improve their reliability.

To understand the possible impact of contaminated fluxes on EPHIN-GOES SEP comparison, it is important to note how these 17 events are distributed across the S-scales. Below is a summary of potentially contaminated events within each S-scale class and their start times:
\begin{itemize}
    \item S1 events (1 out of 40 events): 8/16/01 1:35.
    \item S2 events (5 out of 26 events): 11/6/97 14:40, 5/2/98 16:15, 11/24/00 14:10, 4/18/01 3:40, 11/7/04 18:30.
    \item S3 events (5 out of 11 events): 4/20/98 15:30, 4/3/01 1:00, 1/16/05 1:40, 3/7/12 7:25, 5/22/13 14:45.
    \item S4 events (6 out of 6 events): 7/14/00 11:10, 11/8/00 23:50, 9/24/01 13:35, 11/4/01 17:50, 11/23/01 0:30, 10/28/03 12:25.
\end{itemize}
It is clear that the events experiencing possible contamination constitute a larger fraction of the SEP events as the S-scale increases. Even for the S2 category events, the fraction of potentially-contaminated events is $\sim$20\%, which can possibly explain the decreases in EPHIN-to-GOES peak flux and fluence ratios. Notably, all events in the S4 category (for which GOES has detected approximately $\times$6 peak flux and $\times$3 fluence than EPHIN) experience contamination signatures. This suggests that most of the SEP events observed by GOES, and especially those with $\geq$100\,pfu fluxes of 10-50\,MeV protons, should be interpreted with caution.

\section{Summary and Conclusions}

This study investigated 83 10 - 50\,MeV SPEs detected at L1 by SOHO/EPHIN and at GEO by NOAA’s GOES series during SC 23~\ and most of SC 24. We compare median SPE characteristics detected at these locations and discuss the instrumental and environmental influences on the results, including the correlation of peak flux and fluence ratios with environmental properties (such as the GOES spacecraft location within the magnetosphere, geomagnetic indices, and solar wind properties). The comparison is performed separately for the events with the peak fluxes of 10 - 50\,MeV protons according to GOES measurements of 10$^{1}$-10$^{2}$\,pfu, 10$^{2}$-10$^{3}$\,pfu, 10$^{3}$-10$^{4}$\,pfu, and $>$10$^{4}$\,pfu (which we note as S1, S2, S3, and S4 events in this paper, analogous to SWPC NOAA S-scales). Below is the summary of our findings:

\begin{itemize}
    \item \textbf{Event statistics.} We identified 83 distinct SPEs where 10 – 50\,MeV protons exceeded the 10 pfu threshold in both SOHO/EPHIN and GOES measurements. Additionally, 46 SPEs were detected exclusively by GOES with no EPHIN counterparts, while 96 events were observed by EPHIN without corresponding GOES detections. In both cases, the discrepancies mainly involved weak events with peak fluxes near the 10 pfu threshold, which were excluded from this analysis. Figure~\ref{fig:offset_num} highlights that the MAD values are substantial for many S1–S3 cases, often exceeding the medians themselves for certain timing offsets. This emphasizes that variability between instruments is not negligible and must be carefully accounted for when interpreting weaker SEP events.
    
    \item \textbf{S1 events.} We considered 40 S1 events. On average, SOHO/EPHIN detected earlier onset times ($\Delta t = -20.0 \pm 50.0$\,min), peak times ($\Delta t = -1.0 \pm 1.42$\,hr), event end ($\Delta t = -1.08 \pm 2.21$\,hr), and longer events ($\Delta t = -0.17 \pm 3.54$\,hr) compared to GOES. The EPHIN-to-GOES peak flux and fluence ratios were 1.11$\pm$0.29 and 1.06$\pm$0.45, respectively. These discrepancies suggest that magnetospheric transport or deflection effects play a significant role in shaping weaker SPE measurements at GEO. 
    
    \item \textbf{S2 events.} We examined 26 S2 events, which generally exhibited consistent characteristics across both instruments. None of the investigated properties differed significantly (see Figure~\ref{fig:offset_num}). Among possible explanations of why these events do not experience the same effects as S1 are (1) the higher average particle energies of S2 events, making them less sensitive to magnetospheric effects, and (2) contamination of GOES 10 - 50 MeV fluxes by higher-energy protons, leading to artificially elevated flux measurements.
    
    \item \textbf{S3 and S4 events.} We analyzed 11 S3 and 6 S4 events. For S3, timing was broadly consistent, with significant differences only in event end times, where EPHIN events terminated earlier ($\Delta t = -3.67 \pm 3.33$\,hr). By contrast, GOES detected S4 events slightly earlier ($+2.5 \pm 5.0$\,min) and ended them later ($+0.04 \pm 3.33$\,hr). Both S3 and S4 showed substantially lower EPHIN-to-GOES ratios: $0.84 \pm 0.21$ and $0.16 \pm 0.03$ for peak flux, and $0.75 \pm 0.17$ and $0.29 \pm 0.07$ for fluence, respectively. Our qualitative review indicated contamination signatures in 17 of the 83 analyzed SPEs.

    \item \textbf{Correlations with geomagnetic indices (\textit{Ap, Kp, AE, Dst, solar wind dynamic pressure}).} We examined correlations between the EPHIN-to-GOES peak flux ratios and solar wind and geomagnetic parameters, evaluated separately for each S-category. Overall, only a few statistically significant relationships were identified. The most notable case is for S4 events, which show strong correlation coefficients with \textit{Ap} ($r = -0.73$ and $-0.83$; $p = 0.04$ – $0.06$) and with \textit{Dst} ($r = 0.87$ and $0.94$; $p = 0.02$ – $0.005$). Some weaker but still statistically significant correlations were also observed for S2 and S3 events. However, the correlations are inconsistent across indices and should therefore be interpreted with caution. These findings suggest that, aside from an overall average influence, magnetospheric conditions play only a limited role in modulating GOES 10 – 50\,MeV proton fluxes.
    
    \item \textbf{Other notes.} We also found no evidence of correlations between GOES orbital position or local magnetic pressure with EPHIN-to-GOES flux ratios.   
\end{itemize} 

While our initial expectation was that some differences in particle detections between EPHIN and GOES would primarily arise from transport effects, our analysis shows that this is only part of the picture. We find that EPHIN and GOES measurements are generally consistent for S2 10 – 50\,MeV SPEs, with all key properties mostly agreeing. However, the S1 events demonstrate a delayed start time in GOES along with the weaker peak fluxes and durations, indicating magnetospheric effects. In contrast, S3 and S4 events show signatures of significant contamination of 10 - 50\,MeV channels in GOES. In all three scenarios, the implications of these results depend strongly on their application to prediction models. For example, a median onset-time difference of $\sim$10 minutes is tolerable for day-scale forecasts, but becomes important for short-term warning systems operating on $\sim$30-minute timescales. Similarly, peak and end-time uncertainties of $\sim$60 minutes impose constraints on precise forecasting. Cases of potentially contaminated GOES data further complicate interpretation by reversing onset relationships, inflating peak fluxes and fluences, or extending event durations. This underscores the need to validate GOES observations against independent references such as EPHIN and highlights the inherent uncertainties in datasets commonly used for SEP analysis, which likely limit the accuracy of current forecasting efforts. Many of the observed discrepancies arise from factors beyond the environment through which SEPs propagate, and we find no significant correlations between instrument position, local magnetic pressure, or several geomagnetic parameters. Advancing our understanding of the factors driving flux variations remains essential for improving prediction models, which still cannot fully capture the combined effects of particle transport and magnetospheric interactions on SEP trajectories. The lack of strong correlations between particle flux ratios and geomagnetic or magnetospheric indices suggests that these responses play only a minor role. Moving forward, refining correction methods for GOES flux contamination and exploring the processes that shape particle behavior between L1 and GEO will be important steps toward more accurate SPE analyses. Such efforts will improve forecast reliability and strengthen the foundation for future space exploration missions.

\acknowledgments
We thank Bernd Heber and Arik Posner for their insights and suggestions on adjusting EPHIN data for analysis.
\bibliography{agusample}

@ARTICLE{ephin_noise_kuhl,
       author = {{K{\"u}hl}, P. and {Heber}, B.},
        title = "{Revising More Than 20 Years of EPHIN Ion Flux Data{\textemdash}A New Data Product for Space Weather Applications}",
      journal = {Space Weather},
         year = 2019,
        month = jan,
       volume = {17},
       number = {1},
        pages = {84-98},
          doi = {10.1029/2018SW002114},
       adsurl = {https://ui.adsabs.harvard.edu/abs/2019SpWea..17...84K}
}

@ARTICLE{costep_muller,
       author = {{M{\"u}ller-Mellin}, R. and {Kunow}, H. and {Flei{\ss}ner}, V. and {Pehlke}, E. and {Rode}, E. and {R{\"o}schmann}, N. and {Scharmberg}, C. and {Sierks}, H. and {Rusznyak}, P. and {Mckenna-Lawlor}, S. and {Elendt}, I. and {Sequeiros}, J. and {Meziat}, D. and {Sanchez}, S. and {Medina}, J. and {del Peral}, L. and {Witte}, M. and {Marsden}, R. and {Henrion}, J.},
        title = "{COSTEP - Comprehensive Suprathermal and Energetic Particle Analyser}",
      journal = {\solphys},
         year = 1995,
        month = dec,
       volume = {162},
       number = {1-2},
        pages = {483-504},
          doi = {10.1007/BF00733437},
       adsurl = {https://ui.adsabs.harvard.edu/abs/1995SoPh..162..483M},}

@ARTICLE{me1,
       author = {{Ali}, Aatiya and {Sadykov}, Viacheslav and {Kosovichev}, Alexander and {Kitiashvili}, Irina N. and {Oria}, Vincent and {Nita}, Gelu M. and {Illarionov}, Egor and {O'Keefe}, Patrick M. and {Francis}, Fraila and {Chong}, Chun-Jie and {Kosovich}, Paul and {Marroquin}, Russell D.},
        title = "{Predicting Solar Proton Events of Solar Cycles 22-24 Using GOES Proton and Soft-X-Ray Flux Features}",
      journal = {\apjs},
         year = 2024,
        month = jan,
       volume = {270},
       number = {1},
          eid = {15},
        pages = {15},
          doi = {10.3847/1538-4365/ad0a6c},
archivePrefix = {arXiv},
       eprint = {2303.05446},
 primaryClass = {astro-ph.SR},
       adsurl = {https://ui.adsabs.harvard.edu/abs/2024ApJS..270...15A}
}

@ARTICLE{sep_energies_anastasiadis,
       author = {{Anastasiadis}, Anastasios and {Lario}, David and {Papaioannou}, Athanasios and {Kouloumvakos}, Athanasios and {Vourlidas}, Angelos},
        title = "{Solar energetic particles in the inner heliosphere: status and open questions}",
      journal = {Philosophical Transactions of the Royal Society of London Series A},
         year = 2019,
        month = jul,
       volume = {377},
       number = {2148},
        pages = {20180100},
          doi = {10.1098/rsta.2018.0100},
       adsurl = {https://ui.adsabs.harvard.edu/abs/2019RSPTA.37780100A}
}

@article{radiation_dandouras,
  author    = {Iannis Dandouras and Elias Roussos},
  title     = {High-energy particle observations from the Moon},
  journal   = {Philosophical Transactions of the Royal Society A},
  volume    = {382},
  number    = {2251},
  pages     = {20230311},
  year      = {2024},
  doi       = {10.1098/rsta.2023.0311},
  url       = {https://royalsocietypublishing.org/doi/10.1098/rsta.2023.0311}
}

@article{main_moon_liuzzo,
  author    = {Lucas Liuzzo and Andrew R. Poppe and Christina O. Lee and Shaosui Xu and Vassilis Angelopoulos},
  title     = {Unrestricted solar energetic particle access to the Moon while within the terrestrial magnetotail},
  journal   = {Geophysical Research Letters},
  volume    = {50},
  pages     = {e2023GL103990},
  year      = {2023},
  doi       = {10.1029/2023GL103990},
  url       = {https://doi.org/10.1029/2023GL103990}
}

@article{exploration_hu,
  author    = {Shaowen Hu},
  title     = {Solar Particle Events and Radiation Exposure in Space},
  journal   = {NASA Technical Report},
  year      = {2017},
  url       = {https://three.jsc.nasa.gov/articles/Hu-SPEs.pdf}
}

@article{instrument_jiggens,
  author    = {Piers Jiggens and Marc-Andre Chavy-Macdonald and Giovanni Santin and Alessandra Menicucci and Hugh Evans and Alain Hilgers},
  title     = {The magnitude and effects of extreme solar particle events},
  journal   = {J. Space Weather Space Clim.},
  volume    = {4},
  pages     = {A20},
  year      = {2014},
  doi       = {10.1051/swsc/2014017},
  url       = {https://doi.org/10.1051/swsc/2014017}
}

@article{shelter_martha,
author = {Clowdsley, Martha and Nealy, John and Wilson, John and Anderson, Brooke and Anderson, Mark and Krizan, Shawn},
year = {2005},
month = {02},
pages = {},
title = {Radiation Protection for Lunar Mission Scenarios},
doi = {10.2514/6.2005-6652},
journal = {Aerospace Research Central}
}

@article{shelter_pham,
  author    = {Tai T. Pham and Mohamed S. El-Genk},
  title     = {Dose estimates in a lunar shelter with regolith shielding},
  journal   = {Acta Astronautica},
  volume    = {64},
  pages     = {697--713},
  year      = {2009},
  doi       = {10.1016/j.actaastro.2008.12.002},
  url       = {https://doi.org/10.1016/j.actaastro.2008.12.002}
}

@ARTICLE{pl_example_rotti,
       author = {{Rotti}, Sumanth and {Aydin}, Berkay and {Georgoulis}, Manolis K. and {Martens}, Petrus C.},
        title = "{Integrated Geostationary Solar Energetic Particle Events Catalog: GSEP}",
      journal = {\apjs},
         year = 2022,
        month = sep,
       volume = {262},
       number = {1},
          eid = {29},
        pages = {29},
          doi = {10.3847/1538-4365/ac87ac},
archivePrefix = {arXiv},
       eprint = {2204.12021},
 primaryClass = {astro-ph.SR},
       adsurl = {https://ui.adsabs.harvard.edu/abs/2022ApJS..262...29R}
}

@ARTICLE{pl_example_aminalragia,
       author = {{Aminalragia-Giamini}, Sigiava and {Raptis}, Savvas and {Anastasiadis}, Anastasios and {Tsigkanos}, Antonis and {Sandberg}, Ingmar and {Papaioannou}, Athanasios and {Papadimitriou}, Constantinos and {Jiggens}, Piers and {Aran}, Angels and {Daglis}, Ioannis A.},
        title = "{Solar Energetic Particle Event occurrence prediction using Solar Flare Soft X-ray measurements and Machine Learning}",
      journal = {Journal of Space Weather and Space Climate},
         year = 2021,
        month = nov,
       volume = {11},
          eid = {59},
        pages = {59},
          doi = {10.1051/swsc/2021043},
       adsurl = {https://ui.adsabs.harvard.edu/abs/2021JSWSC..11...59A}
}

@ARTICLE{sickness_lee,
       author = {{Lee}, C.~O. and {Jakosky}, B.~M. and {Luhmann}, J.~G. and {Brain}, D.~A. and {Mays}, M.~L. and {Hassler}, D.~M. and {Holmstr{\"o}m}, M. and {Larson}, D.~E. and {Mitchell}, D.~L. and {Mazelle}, C. and {Halekas}, J.~S.},
        title = "{Observations and Impacts of the 10 September 2017 Solar Events at Mars: An Overview and Synthesis of the Initial Results}",
      journal = {\grl},
         year = 2018,
        month = sep,
       volume = {45},
       number = {17},
        pages = {8871-8885},
          doi = {10.1029/2018GL079162},
       adsurl = {https://ui.adsabs.harvard.edu/abs/2018GeoRL..45.8871L}
}

@article{health_risk_onorato,
  author    = {Giada Onorato and Elia Di Schiavi and Ferdinando Di Cunto},
  title     = {Understanding the Effects of Deep Space Radiation on Nervous System: The Role of Genetically Tractable Experimental Models},
  journal   = {Frontiers in Physics},
  volume    = {8},
  pages     = {362},
  year      = {2020},
  doi       = {10.3389/fphy.2020.00362},
  url       = {https://www.frontiersin.org/articles/10.3389/fphy.2020.00362/full}
}

@ARTICLE{ephin_use1_laurenza,
       author = {{Laurenza}, Monica and {Stumpo}, Mirko and {Zucca}, Pietro and {Mancini}, Mattia and {Benella}, Simone and {Clark}, Liam and {Alberti}, Tommaso and {Marcucci}, Maria Federica},
        title = "{Upgrades of the ESPERTA forecast tool for solar proton events}",
      journal = {Journal of Space Weather and Space Climate},
         year = 2024,
        month = mar,
       volume = {14},
          eid = {8},
        pages = {8},
          doi = {10.1051/swsc/2024007},
       adsurl = {https://ui.adsabs.harvard.edu/abs/2024JSWSC..14....8L}
}

@article{review_whitman,
    title = {Review of Solar Energetic Particle Prediction Models},
    journal = {Advances in Space Research},
    volume = {72},
    number = {12},
    pages = {5161-5242},
    year = {2023},
    note = {COSPAR Space Weather Roadmap 2022-2024: Scientific Research and Applications},
    issn = {0273-1177},
    doi = {https://doi.org/10.1016/j.asr.2022.08.006},
    url = {https://www.sciencedirect.com/science/article/pii/S0273117722007244},
    author = {Kathryn Whitman and Ricky Egeland and Ian G. Richardson and Clayton Allison and Philip Quinn and Janet Barzilla and Irina Kitiashvili and Viacheslav Sadykov and Hazel M. Bain and Mark Dierckxsens and M. Leila Mays and Tilaye Tadesse and Kerry T. Lee and Edward Semones and Janet G. Luhmann and Marlon Núñez and Stephen M. White and Stephen W. Kahler and Alan G. Ling and Don F. Smart and Margaret A. Shea and Valeriy Tenishev and Soukaina F. Boubrahimi and Berkay Aydin and Petrus Martens and Rafal Angryk and Michael S. Marsh and Silvia Dalla and Norma Crosby and Nathan A. Schwadron and Kamen Kozarev and Matthew Gorby and Matthew A. Young and Monica Laurenza and Edward W. Cliver and Tommaso Alberti and Mirko Stumpo and Simone Benella and Athanasios Papaioannou and Anastasios Anastasiadis and Ingmar Sandberg and Manolis K. Georgoulis and Anli Ji and Dustin Kempton and Chetraj Pandey and Gang Li and Junxiang Hu and Gary P. Zank and Eleni Lavasa and Giorgos Giannopoulos and David Falconer and Yash Kadadi and Ian Fernandes and Maher A. Dayeh and Andrés Muñoz-Jaramillo and Subhamoy Chatterjee and Kimberly D. Moreland and Igor V. Sokolov and Ilia I. Roussev and Aleksandre Taktakishvili and Frederic Effenberger and Tamas Gombosi and Zhenguang Huang and Lulu Zhao and Nicolas Wijsen and Angels Aran and Stefaan Poedts and Athanasios Kouloumvakos and Miikka Paassilta and Rami Vainio and Anatoly Belov and Eugenia A. Eroshenko and Maria A. Abunina and Artem A. Abunin and Christopher C. Balch and Olga Malandraki and Michalis Karavolos and Bernd Heber and Johannes Labrenz and Patrick Kühl and Alexander G. Kosovichev and Vincent Oria and Gelu M. Nita and Egor Illarionov and Patrick M. O’Keefe and Yucheng Jiang and Sheldon H. Fereira and Aatiya Ali and Evangelos Paouris and Sigiava Aminalragia-Giamini and Piers Jiggens and Meng Jin and Christina O. Lee and Erika Palmerio and Alessandro Bruno and Spiridon Kasapis and Xiantong Wang and Yang Chen and Blai Sanahuja and David Lario and Carla Jacobs and Du Toit Strauss and Ruhann Steyn and Jabus {van den Berg} and Bill Swalwell and Charlotte Waterfall and Mohamed Nedal and Rositsa Miteva and Momchil Dechev and Pietro Zucca and Alec Engell and Brianna Maze and Harold Farmer and Thuha Kerber and Ben Barnett and Jeremy Loomis and Nathan Grey and Barbara J. Thompson and Jon A. Linker and Ronald M. Caplan and Cooper Downs and Tibor Török and Roberto Lionello and Viacheslav Titov and Ming Zhang and Pouya Hosseinzadeh}
}

@ARTICLE{goes_contamination_posner,
       author = {{Posner}, Arik},
        title = "{Up to 1-hour forecasting of radiation hazards from solar energetic ion events with relativistic electrons}",
      journal = {Space Weather},
         year = 2007,
        month = may,
       volume = {5},
       number = {5},
          eid = {05001},
        pages = {05001},
          doi = {10.1029/2006SW000268},
       adsurl = {https://ui.adsabs.harvard.edu/abs/2007SpWea...5.5001P}
}

@techreport{ephin_doc,
  author       = {Patrick Kuehl},
  title        = {Documentation for COSTEP/EPHIN Level 3 Data (Version 20220201)},
  year         = {2022},
  institution  = {Ulysses/COSTEP Project, University of Kiel},
  url          = {http://ulysses.physik.uni-kiel.de/costep/level3/l3i/DOCUMENTATION-COSTEP-EPHIN-L3-20220201.pdf},
  note         = {Accessed: 2024-12-02}
}

@ARTICLE{ephin_eg1_kuhl,
       author = {{K{\"u}hl}, Patrick and {Heber}, Bernd and {G{\'o}mez-Herrero}, Ra{\'u}l and {Malandraki}, Olga and {Posner}, Arik and {Sierks}, Holger},
        title = "{The Electron Proton Helium INstrument as an example for a Space Weather Radiation Instrument}",
      journal = {Journal of Space Weather and Space Climate},
         year = 2020,
        month = sep,
       volume = {10},
          eid = {53},
        pages = {53},
          doi = {10.1051/swsc/2020056},
archivePrefix = {arXiv},
       eprint = {2010.00864},
 primaryClass = {physics.space-ph},
       adsurl = {https://ui.adsabs.harvard.edu/abs/2020JSWSC..10...53K}
}

@ARTICLE{ephin_eg2_posner,
       author = {{Posner}, A. and {Richardson}, I.~G. and {Strauss}, R.~D. -T.},
        title = "{The ``SEP Clock'': A Discussion of First Proton Arrival Times in Wide-Spread Solar Energetic Particle Events}",
      journal = {\solphys},
         year = 2024,
        month = sep,
       volume = {299},
       number = {9},
          eid = {126},
        pages = {126},
          doi = {10.1007/s11207-024-02350-7},
       adsurl = {https://ui.adsabs.harvard.edu/abs/2024SoPh..299..126P}
}

@article{ephin_eg3_gomez,
title = {SOHO/EPHIN observation of a multiple large solar energetic particles event in November 1997},
journal = {Astroparticle Physics},
volume = {17},
number = {1},
pages = {1-12},
year = {2002},
issn = {0927-6505},
doi = {https://doi.org/10.1016/S0927-6505(01)00136-0},
url = {https://www.sciencedirect.com/science/article/pii/S0927650501001360},
author = {R Gomez-Herrero and M.D Rodriguez-Frias and L {del Peral} and R Müller-Mellin and H Kunow}
}

@ARTICLE{magnetospheric_shielding_kress,
       author = {{Kress}, B.~T. and {Rodriguez}, J.~V. and {Boudouridis}, A. and {Onsager}, T.~G. and {Dichter}, B.~K. and {Galica}, G.~E. and {Tsui}, S.},
        title = "{Observations From NOAA's Newest Solar Proton Sensor}",
      journal = {Space Weather},
         year = 2021,
        month = dec,
       volume = {19},
       number = {12},
          eid = {e02750},
        pages = {e02750},
          doi = {10.1029/2021SW002750},
       adsurl = {https://ui.adsabs.harvard.edu/abs/2021SpWea..1902750K}
}

@ARTICLE{transport_effects_battarbee,
       author = {{Battarbee}, M. and {Guo}, J. and {Dalla}, S. and {Wimmer-Schweingruber}, R. and {Swalwell}, B. and {Lawrence}, D.~J.},
        title = "{Multi-spacecraft observations and transport simulations of solar energetic particles for the May 17th 2012 event}",
      journal = {\aap},
         year = 2018,
        month = may,
       volume = {612},
          eid = {A116},
        pages = {A116},
          doi = {10.1051/0004-6361/201731451},
archivePrefix = {arXiv},
       eprint = {1706.08458},
 primaryClass = {astro-ph.SR},
       adsurl = {https://ui.adsabs.harvard.edu/abs/2018A&A...612A.116B}
}

@article{arrival_direction_filwett,
author = {Filwett, Rachael J. and Jaynes, Allison N. and Baker, Daniel N. and Kanekal, Shrikanth G. and Kress, Brian and Blake, J. Bern},
title = {Solar Energetic Proton Access to the Near-Equatorial Inner Magnetosphere},
journal = {Journal of Geophysical Research: Space Physics},
volume = {125},
number = {6},
pages = {e2019JA027584},
doi = {https://doi.org/10.1029/2019JA027584},
url = {https://agupubs.onlinelibrary.wiley.com/doi/abs/10.1029/2019JA027584},
eprint = {https://agupubs.onlinelibrary.wiley.com/doi/pdf/10.1029/2019JA027584},
note = {e2019JA027584 2019JA027584},
year = {2020}
}

@ARTICLE{particle_energy_obrien,
       author = {{O'Brien}, T.~P. and {Mazur}, J.~E. and {Looper}, M.~D.},
        title = "{Solar Energetic Proton Access to the Magnetosphere During the 10-14 September 2017 Particle Event}",
      journal = {Space Weather},
         year = 2018,
        month = dec,
       volume = {16},
       number = {12},
        pages = {2022-2037},
          doi = {10.1029/2018SW001960},
       adsurl = {https://ui.adsabs.harvard.edu/abs/2018SpWea..16.2022O}
}

@article{bsphere_deflecion_liu,
  author = {Jing Liu and Wenbin Wang and Liying Qian and William Lotko and Alan G. Burns and Kevin Pham and Gang Lu and Stanley C. Solomon and Libo Liu and Weixing Wan and Brian J. Anderson and Anthea Coster and Frederick Wilder},
  title = {Solar flare effects in the Earth's magnetosphere},
  journal = {Nature Physics},
  volume = {17},
  number = {7},
  pages = {807--812},
  year = {2021},
  doi = {10.1038/s41567-021-01203-5}
}

@ARTICLE{Domingo1995SoPh..162....1D,
       author = {{Domingo}, V. and {Fleck}, B. and {Poland}, A.~I.},
        title = "{The SOHO Mission: an Overview}",
      journal = {\solphys},
         year = 1995,
        month = dec,
       volume = {162},
       number = {1-2},
        pages = {1-37},
          doi = {10.1007/BF00733425},
       adsurl = {https://ui.adsabs.harvard.edu/abs/1995SoPh..162....1D}
}

@phdthesis{lsp,
  title={Lightning strike protection and electromagnetic interference shielding for composite structures using metallic submicrofilm and nanofilm},
  author={Bollavaram, Praveen Kumar},
  year={2016},
  school={Wichita State University}
}

@article{electrons,
  author    = {Kirov, B. and Georgieva, K. and Asenovski, S.},
  title     = {Satellite Anomalies and Their Causes},
  journal   = {Sun and Geosphere},
  year      = {2024},
  volume    = {16},
  number    = {1},
  pages     = {10--18},
  doi       = {10.31401/SunGeo.2024.01.02},
  issn      = {2367-8852},
  url       = {https://www.sungeosphere.org/00SGArhiv/SG_v16_No1_2024-pp10-18.pdf},
  note      = {Accessed: 2025-05-09},
  institution = {Space Research and Technology Institute, Bulgarian Academy of Sciences, Sofia, Bulgaria},
  email     = {bkirov@space.bas.bg},
  accepted  = {2024-04-20}
}

@ARTICLE{bruno2018,
       author = {{Bruno}, A. and {Bazilevskaya}, G.~A. and {Boezio}, M. and {Christian}, E.~R. and {de Nolfo}, G.~A. and {Martucci}, M. and {Merge'}, M. and {Mikhailov}, V.~V. and {Munini}, R. and {Richardson}, I.~G. and {Ryan}, J.~M. and {Stochaj}, S. and {Adriani}, O. and {Barbarino}, G.~C. and {Bellotti}, R. and {Bogomolov}, E.~A. and {Bongi}, M. and {Bonvicini}, V. and {Bottai}, S. and {Cafagna}, F. and {Campana}, D. and {Carlson}, P. and {Casolino}, M. and {Castellini}, G. and {De Santis}, C. and {Di Felice}, V. and {Galper}, A.~M. and {Karelin}, A.~V. and {Koldashov}, S.~V. and {Koldobskiy}, S. and {Krutkov}, S.~Y. and {Kvashnin}, A.~N. and {Leonov}, A. and {Malakhov}, V. and {Marcelli}, L. and {Mayorov}, A.~G. and {Menn}, W. and {Mocchiutti}, E. and {Monaco}, A. and {Mori}, N. and {Osteria}, G. and {Panico}, B. and {Papini}, P. and {Pearce}, M. and {Picozza}, P. and {Ricci}, M. and {Ricciarini}, S.~B. and {Simon}, M. and {Sparvoli}, R. and {Spillantini}, P. and {Stozhkov}, Y.~I. and {Vacchi}, A. and {Vannuccini}, E. and {Vasilyev}, G.~I. and {Voronov}, S.~A. and {Yurkin}, Y.~T. and {Zampa}, G. and {Zampa}, N.},
        title = "{Solar Energetic Particle Events Observed by the PAMELA Mission}",
      journal = {\apj},
         year = 2018,
        month = aug,
       volume = {862},
       number = {2},
          eid = {97},
        pages = {97},
          doi = {10.3847/1538-4357/aacc26},
archivePrefix = {arXiv},
       eprint = {1807.10183},
 primaryClass = {astro-ph.SR},
       adsurl = {https://ui.adsabs.harvard.edu/abs/2018ApJ...862...97B}
}

@ARTICLE{connectivity2024,
       author = {{Kouloumvakos}, A. and {Papaioannou}, A. and {Waterfall}, C.~O.~G. and {Dalla}, S. and {Vainio}, R. and {Mason}, G.~M. and {Heber}, B. and {K{\"u}hl}, P. and {Allen}, R.~C. and {Cohen}, C.~M.~S. and {Ho}, G. and {Anastasiadis}, A. and {Rouillard}, A.~P. and {Rodr{\'\i}guez-Pacheco}, J. and {Guo}, J. and {Li}, X. and {H{\"o}rl{\"o}ck}, M. and {Wimmer-Schweingruber}, R.~F.},
        title = "{The multi-spacecraft high-energy solar particle event of 28 October 2021}",
      journal = {\aap},
         year = 2024,
        month = feb,
       volume = {682},
          eid = {A106},
        pages = {A106},
          doi = {10.1051/0004-6361/202346045},
archivePrefix = {arXiv},
       eprint = {2401.05991},
 primaryClass = {astro-ph.SR},
       adsurl = {https://ui.adsabs.harvard.edu/abs/2024A&A...682A.106K}
}

@ARTICLE{kress_cutoff,
       author = {{Kress}, B.~T. and {Hudson}, M.~K. and {Selesnick}, R.~S. and {Mertens}, C.~J. and {Engel}, M.},
        title = "{Modeling geomagnetic cutoffs for space weather applications}",
      journal = {Journal of Geophysical Research (Space Physics)},
         year = 2015,
        month = jul,
       volume = {120},
       number = {7},
        pages = {5694-5702},
          doi = {10.1002/2014JA020899},
       adsurl = {https://ui.adsabs.harvard.edu/abs/2015JGRA..120.5694K}
}

@misc{zenodo,
  author       = {Dröge, H. and Heber, B.},
  title        = {Proton fluxes from the REleASE system from 1995 to 2016 (Version V01) [Data set]},
  year         = {2024},
  publisher    = {Zenodo},
  doi          = {10.5281/zenodo.14191918},
  url          = {https://doi.org/10.5281/zenodo.14191918},
}

@article{coeffs,
  title={Correlation coefficients and their corresponding p-values: misconceptions and examples},
  author={Choi, Bom and Pak, Chunghun},
  journal={Korean Journal of Anesthesiology},
  volume={72},
  number={6},
  pages={575--580},
  year={2019},
  publisher={The Korean Society of Anesthesiologists},
  doi={10.4097/kja.19087},
  pmcid={PMC6375260},
  url={https://doi.org/10.4097/kja.19087}
}

@article{bain_2021,
  author    = {Heather M. Bain and R. A. Steenburgh and T. G. Onsager and E. M. Stitely},
  title     = {A Summary of NOAA Space Weather Prediction Center Proton Event Forecast Performance and Skill},
  journal   = {Space Weather},
  volume    = {19},
  number    = {6},
  pages     = {e2020SW002670},
  year      = {2021},
  doi       = {10.1029/2020SW002670},
  url       = {https://doi.org/10.1029/2020SW002670}
}

@ARTICLE{guo_2023,
       author = {{Guo}, Jingnan and {Wang}, Bingbing and {Whitman}, Kathryn and {Plainaki}, Christina and {Zhao}, Lingling and {Bain}, Hazel M. and {Cohen}, Christina and {Dalla}, Silvia and {Dumbovic}, Mateja and {Janvier}, Miho and {Jun}, Insoo and {Luhmann}, Janet and {Malandraki}, Olga E. and {Mays}, M. Leila and {Rankin}, Jamie S. and {Wang}, Linghua and {Zheng}, Yihua},
        title = "{Particle Radiation Environment in the Heliosphere: Status, limitations and recommendations}",
      journal = {arXiv e-prints},
         year = 2023,
        month = aug,
          eid = {arXiv:2308.11926},
        pages = {arXiv:2308.11926},
          doi = {10.48550/arXiv.2308.11926},
archivePrefix = {arXiv},
       eprint = {2308.11926},
 primaryClass = {physics.space-ph},
       adsurl = {https://ui.adsabs.harvard.edu/abs/2023arXiv230811926G}
}

@ARTICLE{geomag1,
       author = {{Chertok}, I.~M. and {Abunina}, M.~A. and {Abunin}, A.~A. and {Belov}, A.~V. and {Grechnev}, V.~V.},
        title = "{Relationship Between the Magnetic Flux of Solar Eruptions and the Ap Index of Geomagnetic Storms}",
      journal = {\solphys},
         year = 2015,
        month = feb,
       volume = {290},
       number = {2},
        pages = {627-633},
          doi = {10.1007/s11207-014-0618-3},
archivePrefix = {arXiv},
       eprint = {1410.1646},
 primaryClass = {astro-ph.SR},
       adsurl = {https://ui.adsabs.harvard.edu/abs/2015SoPh..290..627C}
}

@ARTICLE{geomag2,
       author = {{Khan}, Parvaiz A. and {Tripathi}, Sharad C. and {Troshichev}, O.~A. and {Waheed}, Malik A. and {Aslam}, A.~M. and {Gwal}, A.~K.},
        title = "{Solar transients disturbing the terrestrial magnetic environment at higher latitudes}",
      journal = {\apss},
         year = 2014,
        month = feb,
       volume = {349},
       number = {2},
        pages = {647-656},
          doi = {10.1007/s10509-013-1661-5},
archivePrefix = {arXiv},
       eprint = {1310.4916},
 primaryClass = {astro-ph.EP},
       adsurl = {https://ui.adsabs.harvard.edu/abs/2014Ap&SS.349..647K}
}

\end{document}